\documentclass[10pt]{iopart}

\usepackage[utf8]{inputenc}
\usepackage{subcaption}
\usepackage[table,xcdraw,dvipsnames]{xcolor}
\usepackage{adjustbox}
\usepackage{amssymb}
\expandafter\let\csname equation*\endcsname\relax

\expandafter\let\csname endequation*\endcsname\relax
\usepackage{amsmath}
\usepackage{upgreek}
\def \be {\begin{equation}}
\def \ee {\end{equation}}
\def \deg  {{$ ^{\circ}$}}

\let\baraccent=\= 
\renewcommand{\=}[1]{\stackrel{#1}{=}} 

\newcommand{\cs}[1]{\langle #1 \rangle}

\usepackage[anticlockwise]{rotating}
\usepackage{float}
\usepackage{epstopdf}
\usepackage{gensymb}
\usepackage{enumitem}
\usepackage{xfrac}
\usepackage[nice]{nicefrac}
\usepackage[small]{titlesec}
\usepackage[toc,page]{appendix}
\usepackage[colorlinks,bookmarks=false,citecolor={black},linkcolor=blue,urlcolor=blue]{hyperref}
\usepackage{caption,todonotes,multirow,amsthm,array,makecell,bm}
\usepackage[backend=biber,style=ieee,citestyle=numeric-comp,sorting=none,natbib=true,isbn=false,url=false,eprint=false, hyperref=true,maxbibnames=1,maxcitenames=1,mincitenames=1,minbibnames=1]{biblatex} 
\usepackage{csquotes}
\DeclareSourcemap{
  \maps[datatype=bibtex, overwrite=true]{
    \map{
      \step[typesource=misc, typetarget=online]
    }
  }
}
\AtEveryBibitem{\clearfield{month}}
\DeclareFieldFormat[article]{volume}{
#1}
\DeclareFieldFormat[article]{pages}{
#1}
\renewbibmacro*{volume+number+eid}{%
  \printfield{volume}%
  \setunit*{\addnbspace}
  \printfield{number}%
  \setunit{\addcomma\space}%
  \printfield{eid}}
\DeclareFieldFormat[article]{number}{\mkbibparens{#1}}

\begin{document}
\title{X-point and divertor filament dynamics from Gas Puff Imaging on TCV}
\author{C. Wüthrich$^{1}$, C. Theiler$^1$, N. Offeddu$^1$, D. Galassi$^1$, D.S. Oliveira$^1$, B.P. Duval$^1$, O. F{\'e}vrier$^1$, T. Golfinopoulos$^2$, W. Han$^2$, E. Marmar$^2$, J.L. Terry$^2$, C.K. Tsui$^3$ and the TCV team$^4$}
\address{ $^1$ {\'E}cole Polytechnique F{\'e}d{\'e}rale de Lausanne (EPFL), Swiss Plasma Center (SPC), CH-1015 Lausanne, Switzerland,\\ 
$^2$MIT Plasma Science and Fusion Center, MA-02139, Cambridge, USA,\\ $^3$Center for Energy Research (CER), University of California San Diego (UCSD), USA,\\ 
$^4$See the author list of H. Reimerdes et al 2022 Nucl. Fusion 62 042018
} 
\ead{curdin.wuethrich@epfl.ch}
\vspace{10pt}
\begin{indented}
\item[]\today
\end{indented}

\begin{abstract}
A new Gas Puff Imaging (GPI) diagnostic has been installed on the 
TCV tokamak, providing two-dimensional insights into Scrape-Off-Layer (SOL) turbulence dynamics above, at and below the magnetic X-point. 
A detailed study in L-mode, attached, lower single-null discharges shows 
that statistical properties 
have little poloidal variations, while
vast differences are present in the 2D behaviour of intermittent filaments. 
Strongly elongated filaments, 
just above the X-point and in the 
divertor far-SOL, show a good consistency in shape and dynamics with field-line tracing from filaments at the outboard midplane, highlighting their connection. 
In the near-SOL of the outer divertor leg, 
short-lived, high frequency and more circular (diameter$\,\sim\!15$ 
sound Larmour radii) filaments 
are observed. These divertor-localised filaments 
appear born radially at the position of maximum density 
and display a radially outward motion with 
velocity $\approx\!400\,$m/s that is comparable to radial velocities of upstream-connected filaments. 
Conversely, in these discharges ($B\times\nabla B$ pointing away from the divertor), these divertor filaments' poloidal velocities 
differ strongly from those of upstream-connected filaments. 
The importance of divertor-localised filaments upon radial transport and profile broadening is explored using filament statistics and in-situ kinetic profile measurements along the divertor leg. 
This provides evidence that these filaments contribute significantly to electron density profile broadening in the divertor. 
\nopagebreak
\end{abstract}

\nopagebreak
\ioptwocol
\section{Introduction}
In a tokamak, the plasma core is magnetically confined on closed magnetic field lines. However, heat and particles are continuously lost into the Scrape-Off Layer (SOL), i.e. the region with open magnetic field-lines, through cross-field transport. In the SOL, this exhaust heat is transported primarily along these open magnetic field lines towards a narrow region of the Plasma Facing Components (PFC), called the divertor target. 
One of the key challenges for next generation fusion devices, such as ITER and DEMO, is in handling this exhaust power \cite{Loarte2007ChapterControl,Pitts2019PhysicsDivertor,Zohm2013OnDEMO}. If unmitigated, peak target heat fluxes are predicted to exceed material limits
, resulting in excessive erosion and melting of the PFCs.

Turbulent transport, predominantly manifested in the form of intermittent structures named filaments \cite{Krasheninnikov2008RecentTurbulence}, is believed to often be the dominant contribution to SOL cross-field transport, especially in the far-SOL \cite{Carralero2015ExperimentalPlasmas}. Convective radial transport associated with filaments, also called blobs because of their shape perpendicular to the magnetic field, has been shown to carry substantial fraction of the power $P_{SOL}$ crossing the Last Closed Flux Surface (LCFS) into the far-SOL \cite{Carralero2018OnContent}. 
Phenomena such as the far-SOL density shoulder along with high levels of potentially harmful plasma interaction with the main chamber first wall have also been associated with filament dynamics \cite{Carralero2017RecentTransport,Tsui2018FilamentaryTokamak,Vianello2020Scrape-offRegimes}. These structures can extend along the magnetic field lines into the divertor \cite{Scotti2018DivertorNSTX-U,OffedduCross-fieldTCV}. 
Their radial velocity, 
SOL penetration distance and the density and heat carried by the filaments, can also significantly determine the 
target heat and particle flux width \cite{Carralero2017RecentTransport}.

Filament generation and dynamics have been extensively studied 
at the low field side outboard midplane of various tokamaks \cite{DIppolito2011ConvectiveExperiment}, such as 
ASDEX-U \cite{Carralero2017RecentTransport}, NSTX \cite{Zweben2017Two-dimensionalNSTX}, Alcator C-Mod \cite{Garcia2018IntermittentPlasmas}, DIII-D \cite{Boedo2001TransportTokamak}, MAST \cite{Kirk2016L-modeImaging} and TCV \cite{Vianello2017ModificationTCV,Tsui2018FilamentaryTokamak,OffedduCross-fieldTCV}. 
Considerably less is, however, known about the properties of filaments around the magnetic X-point and in the divertor. 
A disconnection is expected in the vicinity of the X-point due to flux tube squeezing due to strong flux expansion and magnetic shear \cite{Krasheninnikov2008RecentTurbulence,DIppolito2011ConvectiveExperiment}. 
Radial $\Vec{E}\times\Vec{B}$ shear may act as an additional disconnection mechanism, impacting filaments of larger poloidal sizes depending on the temperature $T_e$ and poloidal field $B_p$ \cite{Nespoli2020AVelocity,OffedduCross-fieldTCV}. 
A better understanding of this is of great importance, as turbulent transport in this region may contribute to 
the spreading of heat and particle radial profiles along the divertor leg \cite{Eich2013ScalingITER}. 
Gas Puff Imaging (GPI) measurements on C-Mod in the region above the X-point showed 
elongated shapes with increased radial velocities compared to the outboard midplane \cite{Terry2009SpatialAlcator-C-Mod}. 
Filament motion around the X-point, both radially and poloidally, has been seen to vary with $B\times\nabla B$ direction on Alcator C-Mod \cite{Terry2017FastC-Mod}. 
Fast visible imaging on MAST \cite{Walkden2017QuiescenceImaging,Harrison2015FilamentaryMAST} and NSTX \cite{Scotti2018DivertorNSTX-U} portrayed a complex picture of filaments entering the divertor forming three different regions of turbulence (PFR, near- and far-SOL common flux). A quiescent region was observed in the near-SOL around the X-point and down to the target plates, together with the existence of small scale filaments intrinsic to the divertor. Those appeared connected to the target and field-aligned up to the X-point, where they poloidally disconnect. 
Radial and poloidal sizes of these divertor filaments of the order of $1$-$3$\,cm were observed and related to resistive-ballooning modes in the bad curvature region of the outer divertor.
The lack of local radial velocity measurements did not, however, allow to draw any conclusions on their contribution to radial particle and heat fluxes. 

A detailed experimental description of the fluctuation properties in the X-point and divertor region improves evaluation of heat and particle redistribution in the divertor, but is also of primary importance for model validation of 
turbulence codes. 
These could, in particular, improve predictive capabilities of the heat flux scale lengths at the target plates that are among the most critical issues for successful operation of ITER and DEMO \cite{Pitts2019PhysicsDivertor}. 
Disentangling the different turbulent mechanisms has recently advanced significantly \cite{Giacomin2020InvestigationSimulations,Giacomin2021Theory-basedDischarges} and real-size, global turbulence simulations of diverted discharges are within reach as demonstrated for MAST \cite{Riva2019Three-dimensionalMeasurements}, ASDEX-U  \cite{Zholobenko2021ElectricLayer} and TCV \cite{Oliveira2022ValidationCase}. Systematic simulation-experimental validation was recently performed for X-point geometries 
on TORPEX \cite{Galassi2022ValidationTORPEX} and TCV \cite{Oliveira2022ValidationCase}. They revealed strong differences in filament dynamics between the codes and experimental measurements  
in the X-point and divertor region, which calls for further studies. 
The detailed nature of radial transport in the divertor is also a key missing component in the performance prediction of long legged, tightly-baffled, divertor scenarios \cite{Umansky2020StudyConfigurations} that show large promise for safe power exhaust in reactors. 

This paper describes the filamentary transport properties at multiple poloidal locations in TCV, focusing on the dynamics around the X-point and along the outer divertor leg of Lower Single-Null (LSN) discharges. A new diagnostic gas puff in the vicinity of the X-point on TCV provides two-dimensional measurements of SOL turbulence properties, focusing upon filament appearance and dynamics, with high spatio-temporal resolution. 
The paper is organised as follows: The diagnostic together with its technical characteristics are described in sec.\,\ref{experiment} with a description of the performed discharges on TCV following in sec.\,\ref{discharges}. In sec.\,\ref{fluctuation}, several statistical moments are discussed across multiple poloidal locations with several diagnostics before analysing in sec.\,\ref{blobshapes} filament shapes and dynamics around the X-point from the X-point GPI. An estimate of their contribution to cross-field transport is presented in sec.\,\ref{fluxcontribution} together with the measurements of 2D density and temperature profiles in the divertor.

\section{X-point GPI system} \label{experiment}

\subsection{Hardware description}
The Tokamak \`a Configuration Variable TCV 
(major radius $R_0$ = 0.88\,m, toroidal field $B_{T} \leq 1.45$\,T) \cite{Reimerdes2022OverviewProgramme} is equipped with many edge turbulence diagnostics. 
Over 180 wall-embedded Langmuir Probes (LP) \cite{Fevrier2018AnalysisVariable,Oliveira2019LangmuirVariable}, a reciprocating probe (RCP) \cite{Tsui2018FilamentaryTokamak} at the outboard midplane and a Reciprocating Divertor Probe Array (RDPA) \cite{Oliveira2021AVariable} allow for 1D and 2D fluctuation measurements of ion saturation current, floating potential and parallel Mach number in the midplane SOL and divertor, respectively. GPI provides 2D turbulence imaging
at 
\begin{figure}[h!b]
    \centerline{\includegraphics[width=8.5cm]{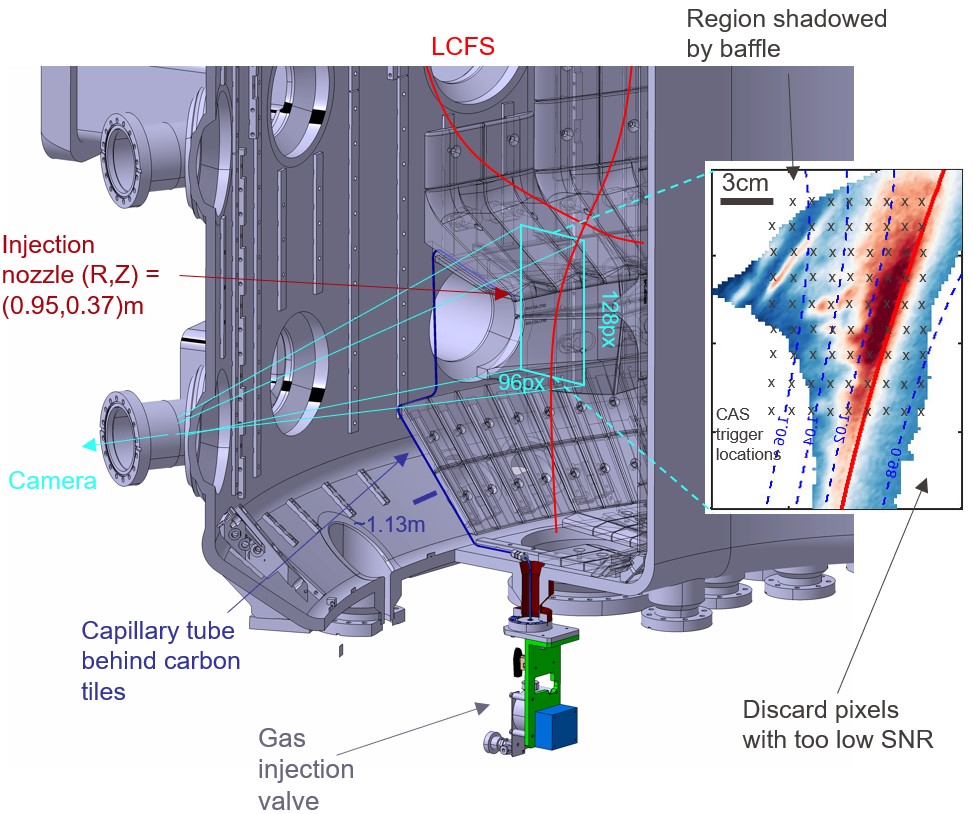}}
    \caption{TCV X-point GPI system. CAD image of the gas injection system 
    in sector 6 of TCV, with the gas puff in the tip of a LFS baffle tile. The X-point GPI viewing geometry and camera
    Field of View (FoV) is illustrated in cyan. 
    A typical plasma separatrix at the toroidal angle of the gas puff is shown in red. 
    A raw image acquired with this setup is shown on the right. Certain parts of the FoV are discarded due to low Signal-To-Noise Ratio (SNR) or baffle shadowing. The crosses indicate the CAS trigger locations.}
    \label{fig:diagnosticcad}
\end{figure}
the outboard midplane of TCV \cite{Han2021SuppressionTCV,OffedduCross-fieldTCV}. However, turbulence imaging in the X-point region was limited up to now.

For this reason, a new GPI system was designed (fig.\,\ref{fig:diagnosticcad}) to obtain insights into the turbulence dynamics across an extended poloidal region around the magnetic X-point of Lower Single Null (LSN) plasmas. Key properties are illustrated here but further diagnostic details will be presented in \cite{OffedduGPITokamak}.
Part of TCV's Plasma Exhaust (PEX) upgrade included the installation of   
baffle tiles on both the Low-Field-Side (LFS) and High-Field-Side (HFS) to separate the main chamber from the divertor \cite{Fasoli2020TCVUpgrades}. Further to its main purpose of studying neutral compression and simplified detachment access \cite{Reimerdes2021InitialDivertor,Fevrier2021DivertorPlasma}, 
a capillary for neutral gas injection was installed in the LFS baffle tip (R\,$\approx\!0.954$\,m, Z\,$\approx\!-0.365$\,m) in the vicinity of the X-point. Gas is injected horizontally into the vessel from a single nozzle with a flow rate of $\approx6\cdot10^{19}$\,at/s and a distance to the separatrix of typically 10-15\,cm, depending on the target plasma geometry. 
The capillary tube (blue in fig.\,\ref{fig:diagnosticcad}) runs behind the graphite tiles to the bottom of the vacuum vessel, where a feedback controlled piezoelectric valve controls the flow rate. 
Helium (He) gas puffs were used in all of the discharges described in this work, though the system also permits for deuterium (D$_2$) injection. 
This was chosen as the background light emission from D$_2$ can be orders of magnitude higher than at the outboard midplane, displaying significant emissivity and reduced toroidal localisation of the light from the GPI injected neutrals \cite{Zweben2017InvitedDevices}. 
\vspace{-0.2cm}
\begin{figure}[!h]
    \centerline{\includegraphics[width=7.5cm]{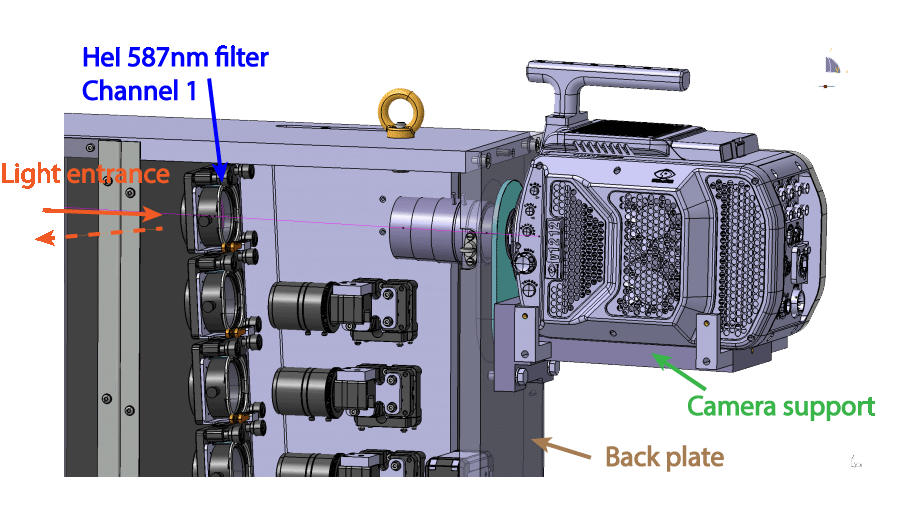}}
    \caption{Schematic illustration of the fast-imaging Phantom camera and its mounting system on MANTIS. It views the first channel without disturbing the other spectral cameras (the orange dashed arrow shows the reflected light passage to the rest of MANTIS).}
    \label{fig:cameracad}
\end{figure}
A magnetically shielded Phantom v2012 high speed camera ($400$\,kHz, exposure time $2\,\upmu$s, resolution of 126$\times$96\,px$^2$) is used.\footnote{\url{https://www.phantomhighspeed.com/products/cameras/ultrahighspeed/v2012}} 
The camera was introduced into the Multispectral Advanced Narrowband Tokamak Imaging System (MANTIS) \cite{Perek2019MANTIS:Plasmas}, as shown in fig.\,\ref{fig:cameracad}, installed on TCV on the lower tangential port of sector 8, with the light relayed from the port through a set of mirrors and relay optics to a HeI 587.5\,nm filter (2\,nm bandwidth, 95\% transmission). This allows for the camera to be farther from TCV, minimising any magnetic field influence while avoiding optical fibres. 

An absolute spectral calibration allows for measurements of the local emissivity.
Each camera pixel collects light from a 1.9$\times$1.9\,mm$^2$ region in the poloidal plane at the injection nozzle. 
The diagnostic Lines-Of-Sight (LOS) are well aligned with the magnetic field in the gas puff plane, with a poloidal and toroidal misalignment of $\approx\,$5\deg and $\approx\,$3\deg, respectively, for discharges with standard helicity ($I_p$ same sign as $B_T$) and geometries as those studied in this work. The toroidal gas cloud extent along the magnetic field lines was estimated 
to be 5-10\,cm, depending on the distance to the nozzle. This results in a limiting resolution of $\leq7$\,mm as estimated using the approach in sec. II.E. in \cite{Zweben2017InvitedDevices}. 
Smearing due to the integration time of $2\,\upmu$s can add up to 2\,mm to this limit for cross-field velocities of 1\,km/s. 
The position calibration of the pixel views in the poloidal plane is performed using calcam raytracing \cite{Silburn2018Calcam} and typically results in an uncertainty of 1 pixel.

\subsection{Signal analysis and relation to density fluctuations}
The brightness $I$ 
measured by the GPI camera detector in combination with an adequate brightness calibration can be related to plasma parameters as 
\cite{Zweben2017InvitedDevices,Churchill2013DevelopmentEdge} 
\begin{equation}
    I = \frac{1}{4\pi}\int_{LOS} n_n
    f\,\, dl \simeq \frac{1}{4\sqrt{\pi\ln{2}}} L\, n_n^{max} n_e^{\alpha} T_e^{\beta}  \label{eq:brightness}
\end{equation}
with $n_n$ the injected neutral density available for excitation at the observed wavelength and $n_e$ and $T_e$ the local electron density and temperature, respectively. $\int_{LOS}dl$ depicts the integral along the optical LOS of each pixel through the neutral gas cloud. For the approximate expression on the right of eq. \ref{eq:brightness}, a Gaussian profile for $n_n$ along the LOS was assumed, with a peak value of $n_n^{max}$ and a half width half maximum $L$, of sufficient narrowness that $n_e$ and $T_e$ may be considered constant. 
$f(n_e,T_e)$ represents the ratio of neutrals in the upper excited state of the radiative transition to the ground state times the rate coefficient of this decay and is commonly rewritten 
as a power law function with exponents $\alpha=\alpha(n_e,T_e)$ and $\beta=\beta(n_e,T_e)$ \cite{Zweben2017InvitedDevices}.
Such a dependence makes the relation between plasma parameters and GPI intensity fluctuations non-linear and complicated. 

A measure of the relative fluctuation amplitude can be obtained by normalising the brightness to the moving average of the signal as

\begin{align} 
S=\frac{I-\cs{I}_{1ms}}{\cs{I}_{1ms}} & \sim \frac{n_n(n_e^{\alpha}T_e^{\beta} - \cs{n_e^{\alpha}T_e^{\beta}})}{n_n \cs{n_e^{\alpha}T_e^{\beta}}} \label{eq:normalizedbright}\\
 & \approx \alpha \frac{n_e-\cs{n_e}}{\cs{n_e}} + \beta \frac{T_e-\cs{T_e}}{\cs{T_e}} \label{eq:normalizedbrightapprox}
\end{align}
where $\cs{I}$ describes the moving average of the pixel-wise brightness over typically 1\,ms. Slow fluctuations of the neutral density, low frequency plasma fluctuations and other effects due to cloud non-uniformity 
are removed in this procedure \cite{Zweben2017InvitedDevices}. $S$ is thus a quantity proportional to the relative density and temperature fluctuations. 
Eq.\,\ref{eq:normalizedbrightapprox} is valid only for small fluctuations relative to the mean values of $n_e$ and $T_e$ (see also \cite{Kube2020ComparisonLayer}). For the investigated plasmas with $n_e \in [0.3,1]\cdot10^{19}\,$m$^{-3}$ and $T_e \in [5,25]$\,eV, one expects $\alpha \in [0.4, 0.6]$ and $\beta \in [0.3,4]$ according to \cite{Zweben2017InvitedDevices}. 
For this parameter range, we have $\alpha(n_e,T_e)\simeq\alpha(n_e)$ and  $\beta(n_e,T_e)\simeq\beta(T_e)$ resulting in an approximately linear dependence.

\subsection{Image processing}
In the diagnostic setup described above, the upper part of the FoV looks at the underside of the LFS baffle, resulting in the shadowed area indicated in the example image on the right of fig.\,\ref{fig:diagnosticcad}. 
The corresponding pixels should not be employed. Regions of the view with a SNR below 12 are also excluded, where the SNR is evaluated as the ratio between the mean intensity during stationary gas injection and the standard deviation $\sigma_{dark}$ before the gas puff, where in the peak signal region, a SNR$\sim60$-$80$ is typically observed.
In the remainder of this work, images recorded on the camera are horizontally mirrored, such that the LFS region is located to the right, as is generally the practice. 

The image series are mostly analysed using Conditional Average Sampling (CAS) \cite{Zweben2017InvitedDevices} with a triggering threshold of $S>2.5\sigma$. A total of 81 trigger locations are distributed over the FoV, as shown in fig.\,\ref{fig:diagnosticcad}. The average radial and poloidal velocities of the filaments at these locations are then extracted from the Centre of Light (CoL) displacement, averaged over 5 CAS frames.

\section{TCV L-mode scenarios} \label{discharges}
\begin{figure}[b]
    \centerline{\includegraphics[width=4.1cm]{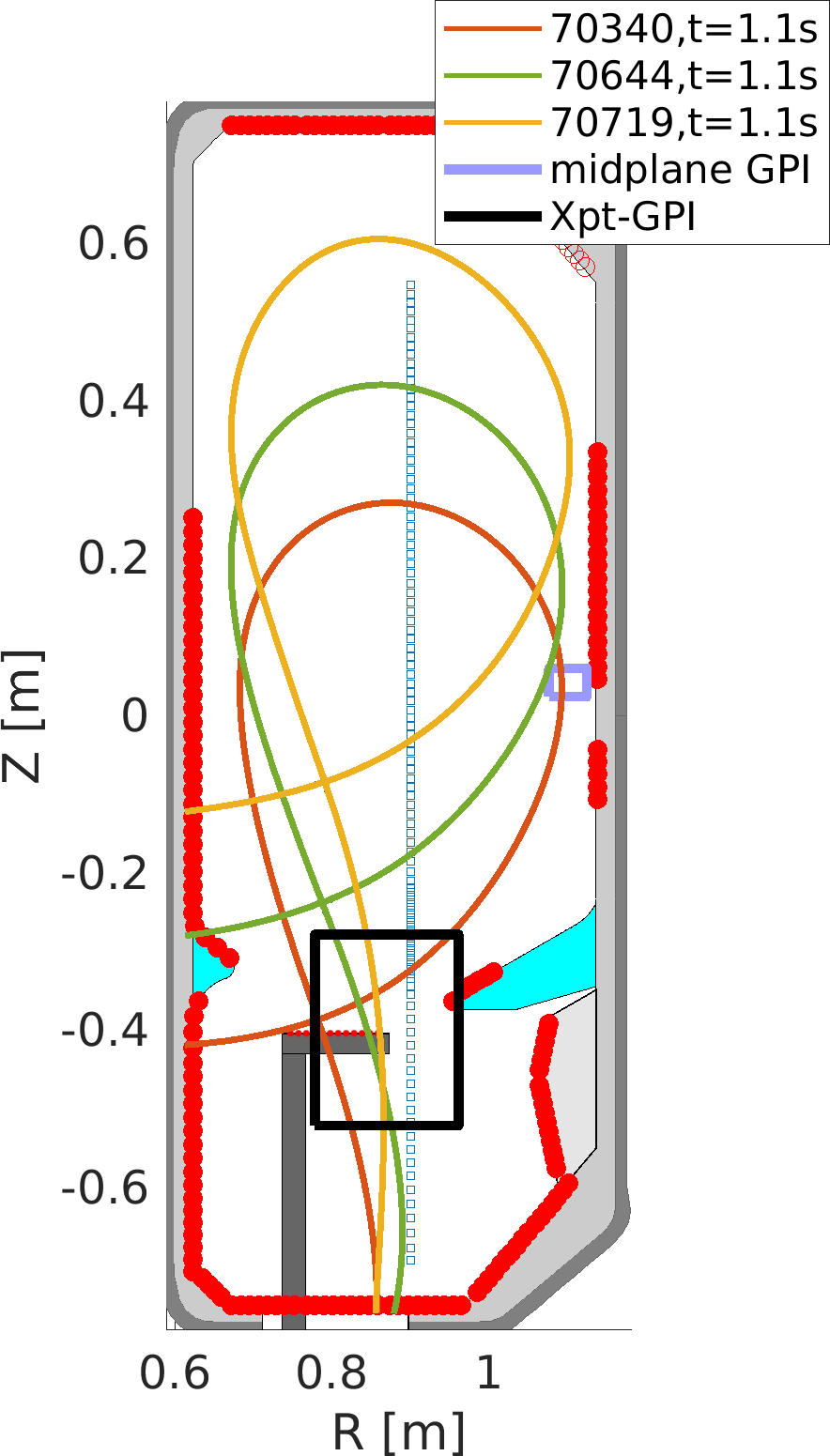}}
    \caption{The three similar, but vertically displaced magnetic LSN geometries considered in this work, together with the edge turbulence diagnostic positions: the X-point and midplane GPI FoV (black and lila rectangle), the RDPA (grey structure) 
    and wall-LPs (red dots). The blue dots indicate the measuring volumes of the Thomson Scattering (TS). The colour coding for GPI data is maintained throughout this paper: red for data obtained from the lower core, green from below the X-point and yellow from the outer divertor leg.}
    \label{fig:liuqegeom}
\end{figure}
This work focuses on TCV lower single null, low density, L-mode configurations ($B_T=1.4$\,T, $I_p=245$\,kA, line average density $\approx\!3.9\cdot10^{19}$\,m$^{-3}$, Greenwald fraction $f_G\!\approx\!0.23$) 
with magnetic geometry as shown in fig.\,\ref{fig:liuqegeom}. The figure also indicates the locations of the used turbulence diagnostics. The discharges are performed in reversed field, i.e. with the ion B$\times \nabla $B direction chosen to reduce H-mode access. The plasma is fuelled from the top into the main chamber to avoid effects from the baffles. An identical plasma was vertically displaced by $\pm 15$\,cm compared to the central case (green), for turbulence measurements with the LFS X-point GPI probing a region above and below the X-point. 
An identical discharge (\#71575) to \#70340 (red geometry in fig.\,\ref{fig:liuqegeom}) but in non-standard helicity ($I_p=-245$\,kA) is added for better alignment with the sight lines of the outboard midplane GPI system. 
The connection length from the outboard midplane to the outer target varies by $\pm 15 \%$ 
across the three geometries in fig. \ref{fig:liuqegeom}. 
Other relevant plasma parameters, such as triangularity, target flux expansion ($f_x \approx 3.5$) and strike point position, were essentially kept constant. 

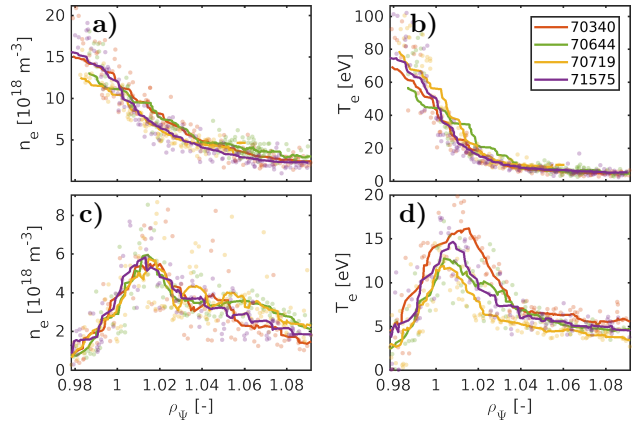
\begin{figure}[!ht]
\begin{picture}(0,160)
         \put(0,0){
    \centerline{\includegraphics[width=.99\linewidth]{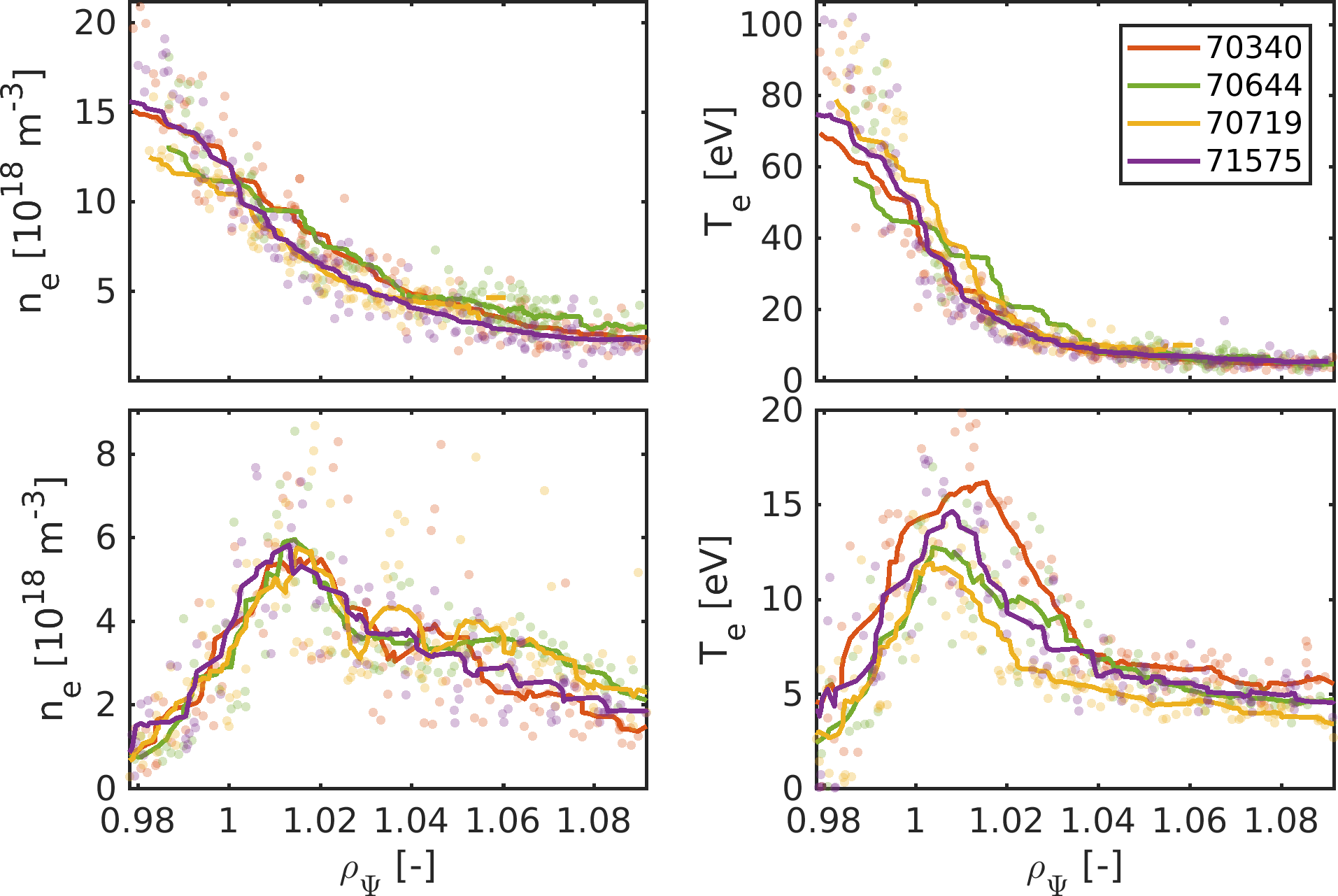}}}
    \put(35,147){\textbf{a)}}\put(155,147){\textbf{b)}}\put(30,75){\textbf{c)}}\put(151,75){\textbf{d)}}
    \end{picture}
    \caption{Kinetic radial plasma profiles from the three equilibria in fig.\,\ref{fig:liuqegeom} of $n_e$ and $T_e$ from TS (a-b) by considering only the chords in the lower half of the vessel. The normalised flux coordinate $\rho_{\Psi}$ is defined as $\sqrt{(\Psi-\Psi_0)/(\Psi_{LCFS}-\Psi_0)}$, with $\Psi_0$ and $\Psi_{LCFS}$ the poloidal magnetic flux on-axis and at the LCFS, respectively. The same quantities are extracted from wall-embedded LPs (c-d) on the outer divertor target.}
    \label{fig:neteprofiles}
\end{figure}

Displacing the plasma is, therefore, taken as a good approximation, to moving the GPI view for 
turbulence measurements. 
In the following, these measurements are super-imposed 
onto the green geometry in fig.\,\ref{fig:liuqegeom} by 
vertical translations. 
This approach is further supported by fig.\,\ref{fig:neteprofiles}, showing the electron density $n_e$ and temperature $T_e$ profiles from TS in the lower half of the core together with the floor LP profiles of the same quantities. Minimal differences are observed both in the upstream SOL and at the floor, indicating a weak impact of the outer divertor leg length on the measurements for these plasma conditions.

A leading theory for filament propagation is the two region model \cite{Myra2006CollisionalityBlobs}. 
Using the $n_e$ and $T_e$ profiles from TS in fig.\,\ref{fig:neteprofiles} and the midplane GPI data, following the procedure in \cite{OffedduCross-fieldTCV}, we conclude these discharges to be in the Resistive X-point regime. In such a regime, it is postulated that the ion polarisation current is the dominant current return path, that occurs around the X-point due to strong filament squeezing. Whether the current closure occurs past the X-point or not depends on the collisionality.

\section{Statistical moments of fluctuations} \label{fluctuation}
Before focusing specifically on filaments, we investigate here the fluctuation characteristics by looking at the first three statistical moments of the measured brightness intensity $I$.

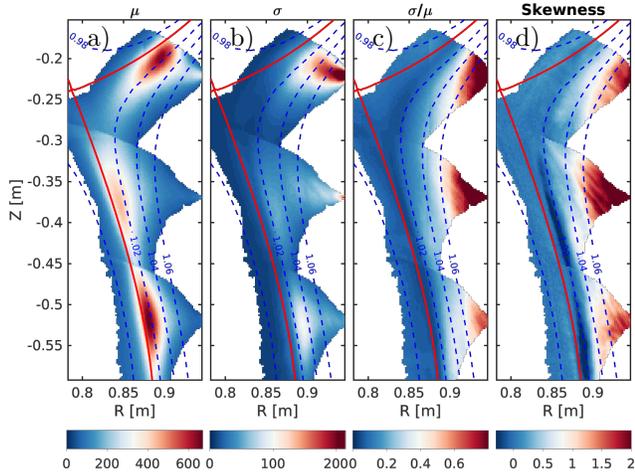
\begin{figure}[t]
    \begin{picture}(0,178)
    \put(0,0){
    \centerline{\includegraphics[width=8.3cm]{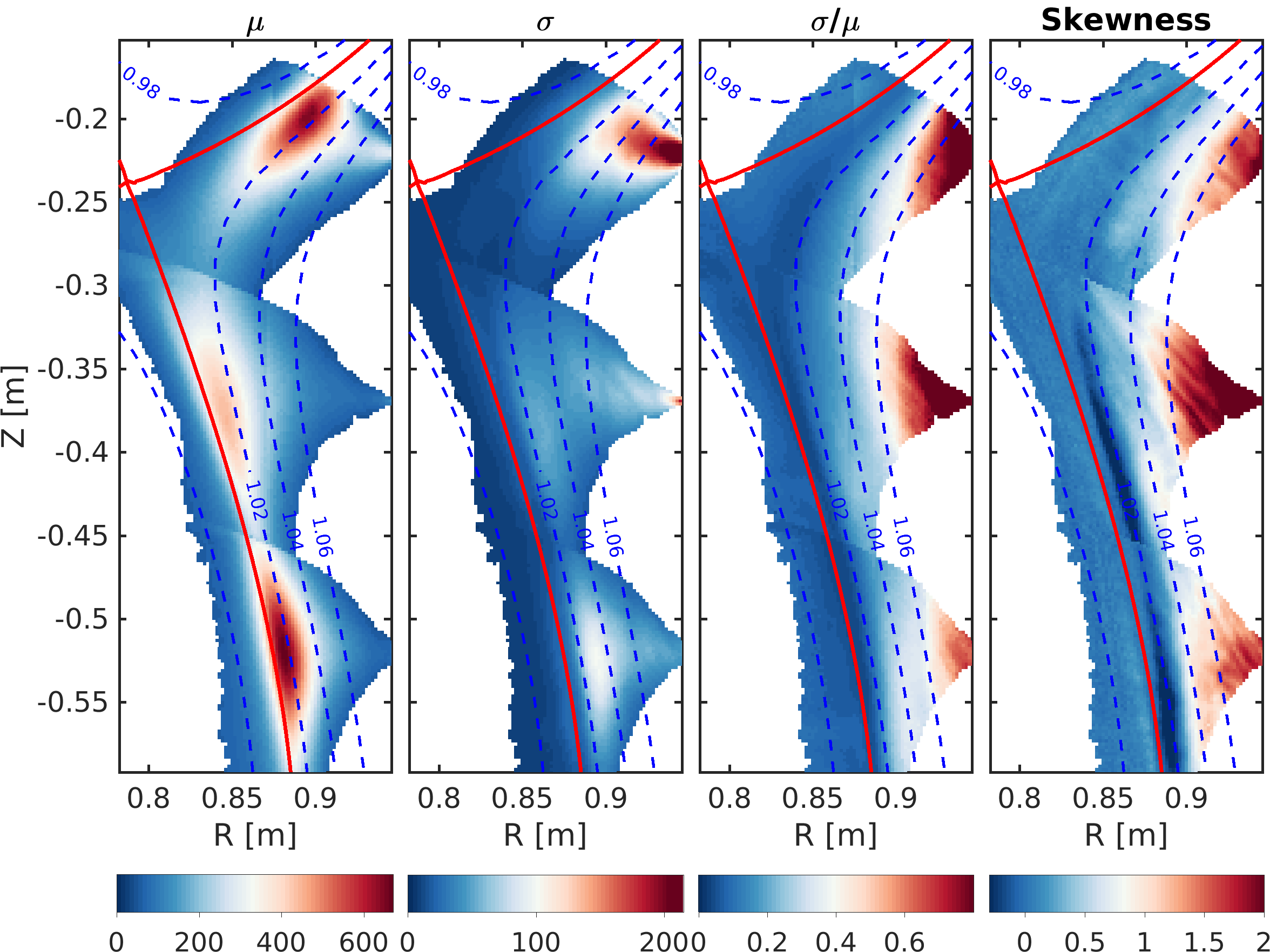}}}
    \put(33,160){a)}\put(87,160){b)}\put(141,160){c)}\put(195,160){d)}
    \end{picture}
    \caption{Statistical moments of GPI brightness (from a to d): mean $\mu$ and standard deviation $\sigma$ in arbitrary brightness units, relative fluctuation level $\sigma/\mu$ and skewness. 
    The blue dashed lines indicate flux surfaces with $\rho_{\Psi}\!=\![0.98,1.02,1.04,1.06]$ and the red line the separatrix. Remarkable uniformity in $\sigma/\mu$ is observed for $\rho_{\Psi}\in [1.01,1.03]$. $\sigma/\mu$ broadens around $\rho_{\Psi}\approx1.06$ at the height of the X-point.}
    \label{fig:stat2dmap}
\end{figure}

The 2D measurement of these quantities in the 
different regions around the X-point, are shown in fig.\,\ref{fig:stat2dmap}. 
The computation is made piece-wise over a moving 4\,ms time window before taking the mean over the full window to average low frequency components. Diagnostic-related noise fluctuations are reduced by subtracting the standard deviation before the He gas puff, $\sigma_{dark}$, from that during the measurement window ($\sigma_{meas}$). Likewise, the background brightness before injection, $\mu_{dark}$, is subtracted and an analysed relative fluctuation level is deduced as $\sigma=\frac{\sqrt{\sigma_{meas}^2-\sigma_{dark}^2}}{\mu-\mu_{dark}}$.

The mean brightness, fig.\,\ref{fig:stat2dmap}a, increases from the injection nozzle towards the region of higher plasma density, reaching a maximum just before the LCFS in all three probed regions. In the divertor leg, the maximum brightness coincides well with the peak electron density according to fig.\,\ref{fig:neteprofiles}c. Beyond the separatrix, the signal decreases rapidly, either due to low plasma density in the PFR or from ionisation of injected neutrals in the plasma core. The standard deviation (fig.\,\ref{fig:stat2dmap}b) shows a similar behaviour.

The relative fluctuation amplitude $\sigma/\mu$ 
(fig.\,\ref{fig:stat2dmap}c) 
shows a remarkable poloidal uniformity over the entire measurement region in the near SOL and throughout the divertor. It increases radially monotonically up to $\sim80\,\%$ at $\rho_{\Psi}\approx 1.06$-$1.08$. At the divertor entrance, however, a transition is observed in the far-SOL, resulting in a broadening of the zone of low fluctuations. 
A decrease of $\approx$\,0.2 is observed for $\rho_{\Psi}\!\approx\!1.06$ from just above to just below the X-point, with no significant further poloidal variation along the divertor leg. 
The region of $1.01\!<\!\rho_{\Psi}\!<\!1.02$ shows the lowest fluctuation level of $\sim\,$10\%, corresponding to the radial location of maximum density
, reminiscent of a quiescent region observed on MAST \cite{Walkden2017QuiescenceImaging}. 
In the PFR and in the core above the magnetic X-point, fluctuations remain low.

The skewness (fig.\,\ref{fig:stat2dmap}d) shows a similar poloidal uniformity. In contrast to $\sigma/\mu$, no significant poloidal variation is observed in the far-SOL above and below the X-point. 
Around the region of minimum $\sigma/\mu$, the skewness becomes close to zero and even negative in the divertor leg. 
A radial increase to $\approx\!1.5$ at $\rho_{\Psi}\!\approx\!1.08$ is observed, indicative of enhanced filamentary activity. 
Striations are observed in the far-SOL skewness, consistent in size and tilt with typically observed elongated filaments in this region, further discussed in section \ref{blobshapes}. Such striations were also previously reported using passive fast imaging analysis on MAST \cite{Walkden2017QuiescenceImaging}. 

\begin{figure}[t]
    \centerline{\includegraphics[width=.9\linewidth]{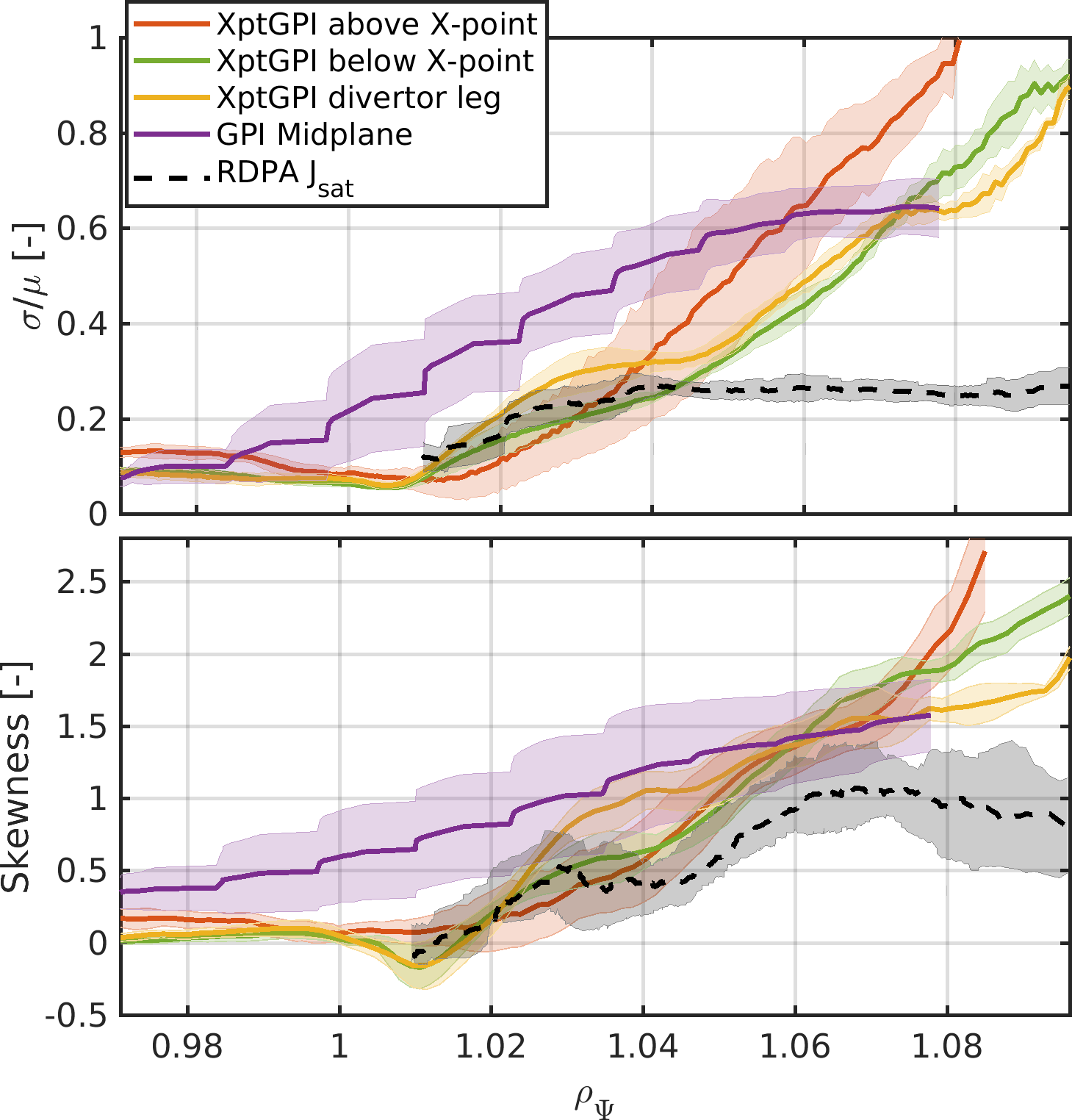}}
    \caption{Radial profiles of rel. fluctuation level (top) and skewness (bottom) for all X-point GPI locations as compared to the outboard midplane. 
    Also shown are the relative $J_{sat}$ fluctuations from RDPA in the outer divertor leg of discharge \#70340, 
    at the same vertical distance to the X-point ($D_{xpt}$) as the GPI data labelled "below X-point".}
    \label{fig:statprofile}
\end{figure}
These fluctuation measurements are now compared to GPI at the outboard midplane. 
In fig.\,\ref{fig:statprofile}, radial profiles of $\sigma/\mu$ and skewness as a function of $\rho_{\Psi}$ 
are shown. In addition, $J_{sat}$
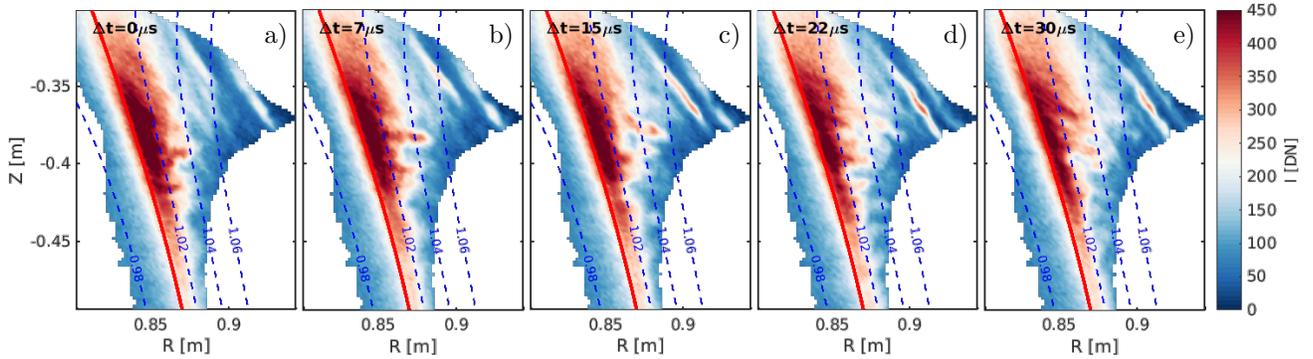
\begin{figure*}[!h]
\begin{picture}(0,135)
         \put(0,0){
    \centerline{\includegraphics[width = \linewidth]{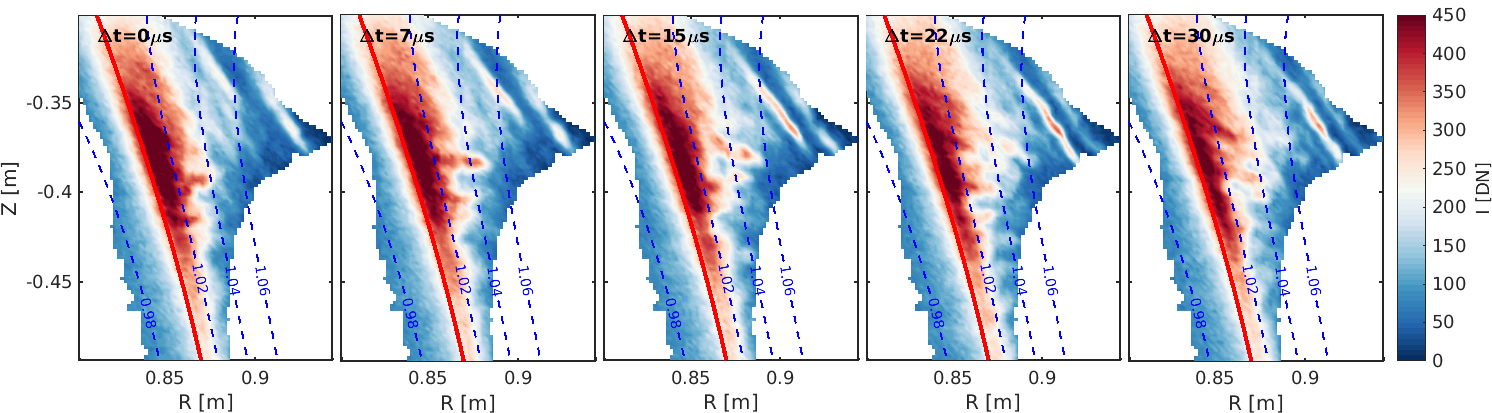}}}
    \put(100,118){a)}\put(185,118){b)}\put(272,118){c)}\put(357,118){d)}\put(444,118){e)}
    \end{picture}%
    \caption{False colour raw image sequence of discharge \#70644 viewing below the X-point in the divertor leg. Every third frame is shown for better visibility of the structures. The red line shows the separatrix and blue dashed lines indicate flux surfaces at $\rho_{\Psi}=[0.98,1.02,1.04,1.06]$, a convention followed throughout this paper.}
    \label{fig:snapshot}
\end{figure*}
fluctuations measured by the RDPA in the divertor volume at a distance of $D_{xpt}\,\approx\,15$\,cm from the X-point 
are displayed. 

This shows well the decrease of $\sigma/\mu$ from above to below the X-point in the far-SOL but also a slight increase around $\rho_{\Psi}\,\approx\,1.03$ in the divertor near-SOL. 
At the outer-midplane, fluctuations are significantly larger, by $\approx0.3$ in the near-SOL. This suggests more turbulence at the transition from open to closed magnetic field lines at the outer midplane.
The skewness is also higher at the outboard midplane compared to the other poloidal regions, at least in the near-SOL for $\rho_{\Psi} \leq 1.04$.

The cross-comparison of RDPA (black curves in fig.\,\ref{fig:statprofile}) and the X-point GPI at the same location (green curve) 
shows good agreement in both statistical quantities for $\rho_{\Psi}\in[1.01,1.05]$. 
Discrepancies are observed in relative fluctuation level radially further out. This could be related to the difference in power dependence on $n_e$ and $T_e$ between the ion saturation current measured by the RDPA and the He line emission intensity from GPI. A comparison between LPs on the floor and GPI at the outer midplane on C-Mod \cite{Kube2020ComparisonLayer} showed similar trends 
with differences of up to 0.3 in $\sigma/\mu$ in the far-SOL. 
The same effect could, to some extent, explain the difference in the far-SOL between the midplane GPI, where neutral D$_2$ is puffed, and the X-point GPI using He injection. The two gases have different $n_e$ and $T_e$ dependencies in line emission intensity (eq.\,\ref{eq:brightness}). Previous experiments with identical discharges measured with the midplane GPI on TCV, once with D$_2$ and once with He, showed agreement in the near-SOL 
and an increase of 0.1 with He in the far-SOL.

\section{Filament properties 
around X-point} \label{blobshapes}
In this section, we discuss 2D shapes, origin and velocities of filaments observed by the X-point GPI.
A series of raw GPI snapshot images over 30$\upmu s$ from below the X-point, displayed in fig.\,\ref{fig:snapshot}, shows many characteristic image features. 
As seen in the mean statistical moment (fig.\,\ref{fig:stat2dmap}a), the brightest region is located at and just outside 
the separatrix. 
Two distinct types of structures appear. Highly elongated structures are observed in the far-SOL at $\rho_{\Psi}>1.04$, moving radially outwards and poloidally in the direction of the target. 
In contrast, a more complex fluctuation pattern is observed closer to the separatrix at $1.01< \rho_{\Psi}<1.05$. Multiple structures, smaller in size, appear along the leg at several positions, seemingly detaching from the brightest area and then moving towards the LFS. To the left of the peak brightness region, in the PFR, no significant fluctuations or filaments are identified all along the FoV. 
Thin, poloidally elongated filaments in the PFR, as documented on C-Mod \cite{Terry2017FastC-Mod} and MAST \cite{Harrison2015FilamentaryMAST}, were not detected here.

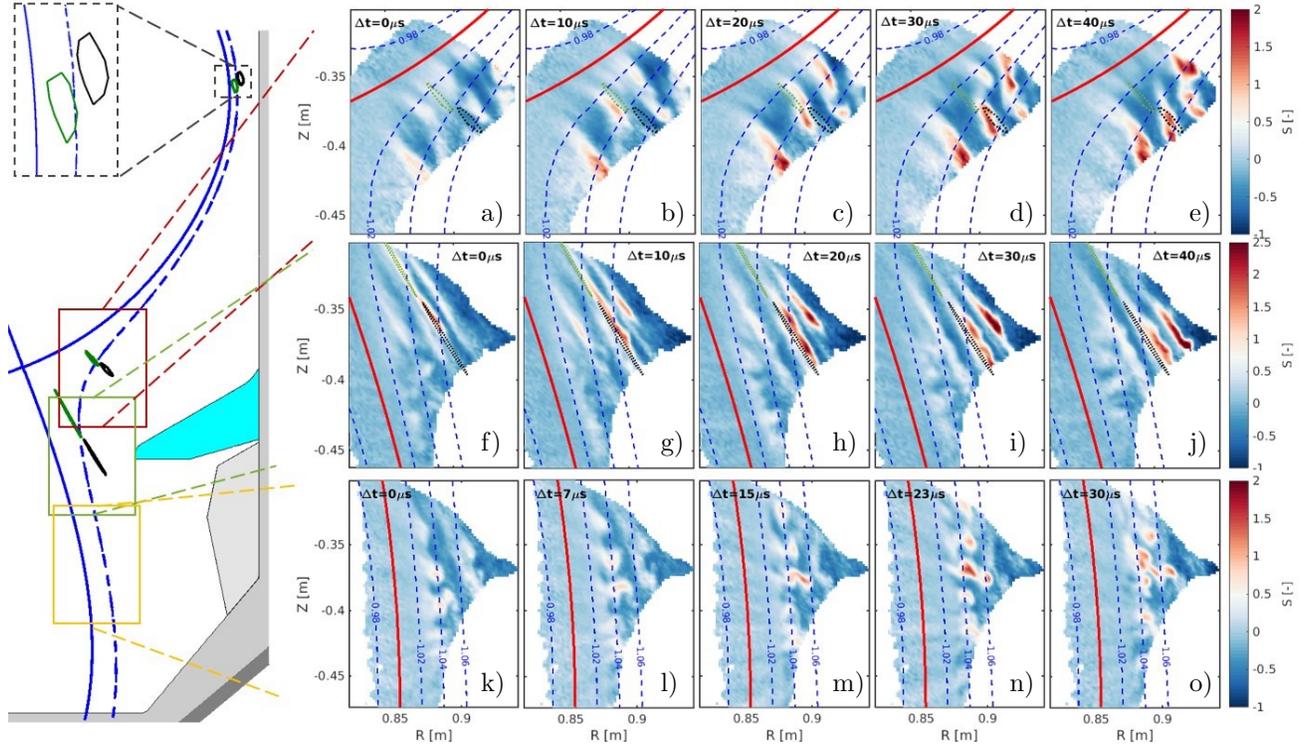
\begin{figure*}[h!t]
    \begin{picture}(0,280)
         \put(0,0){
    \centerline{\includegraphics[width=\linewidth]{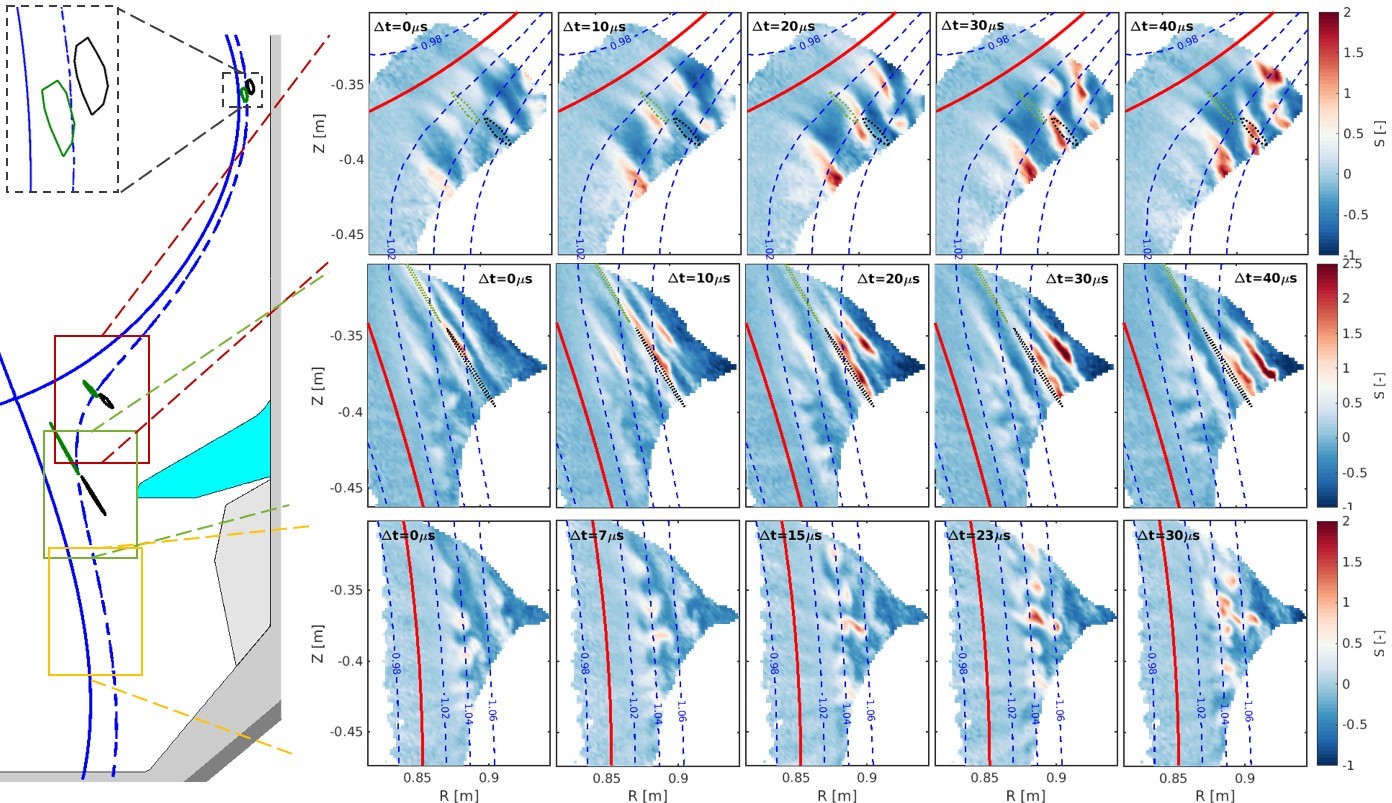}}}
    \put(182,197){a)}\put(250,197){b)}\put(315,197){c)}\put(382,197){d)}\put(450,197){e)}
    \put(182,110){f)}\put(250,110){g)}\put(315,110){h)}\put(382,110){i)}\put(450,110){j)}
    \put(182,19){k)}\put(250,19){l)}\put(315,19){m)}\put(382,19){n)}\put(450,19){o)}
    \end{picture}
    \caption{Image sequences of normalised GPI brightness $S$, above the X-point and in the divertor. In the two lower regions, both elongated filaments in the far-SOL and circular structures in the divertor leg are present. The green and black dotted contours correspond to the half maximum contours from  
    conditionally averaged filaments, with a time separation of $\approx 16\,\upmu$s, detected with the midplane GPI in discharge \#71575. They are traced along magnetic field lines to the X-point regions as shown in the magnetic reconstruction on the left.} 
    \label{fig:snapshots}
\end{figure*}

The snapshot series of the normalised signal $S$ (eq.\,\ref{eq:normalizedbright}) in fig.\,\ref{fig:snapshots} provides further insight. 
Just above the X-point, panels a)-e), clear structures are visible in the SOL starting at $\rho_{\Psi}\approx1.01$, followed by a radial outwards displacement. Regularly, multiple filaments are visible simultaneously within the observation window 
and they often shear apart in the far-SOL around $\rho_{\Psi}\approx1.06$ (d). Stronger poloidal motion is observed closer to the LCFS together with less distinct filament like features. Overall, the pattern above the X-point agrees well with GPI studies on C-Mod \cite{Terry2009SpatialAlcator-C-Mod}.

In the region below the X-point, panels f)-j), as already noted, two regions can be distinguished. The structures in the far-SOL have a large aspect ration and show a near vertical tilting 
with strong fluctuation amplitude. 
Independently, smaller scale and less intense filaments are observed 
from $D_{xpt}\approx\,13$\,cm below the X-point onwards. 
In the bottom window (panels k-o) splitting and complex interactions between filaments can be seen. They are present all along the divertor leg and can travel across the full common flux width. Beyond $\rho_{\Psi}>1.06$, poloidally elongated features are still occasionally observed. 

These TCV divertor observations are 
consistent with the theoretical picture elongated cross-sections of SOL filaments around the X-point region due to increased flux expansion and magnetic shear \cite{Krasheninnikov2008RecentTurbulence}, discussed further in sec.\,\ref{magmap}. The picture shown here also largely mirrors passive fast imaging measurements on MAST 
\cite{Harrison2015FilamentaryMAST,Walkden2017QuiescenceImaging}. 

\subsection{Magnetic mapping and identification of divertor localised filaments} \label{magmap}
Magnetic field line tracing \cite{OffedduCross-fieldTCV} is used to 
identify the filament types in the LFS poloidal locations and their possible upstream connection. Two CAS contours are taken from the midplane GPI (indicated in green and black in the top left of fig.\,\ref{fig:snapshots}) corresponding to a filament displacement over 16\,$\upmu$s. 
These contours are traced along the magnetic field lines down to the LFS outer divertor views. 
The field lines intersect with the GPI views above and below the X-point after $\sim$\,0.5 and 1 toroidal turns, respectively. 

The dimension, tilt and location of the structures just above the X-point and in the divertor far-SOL are well matched by shapes expected from upstream. 
The typical time scale of the filament passage and the displacement direction also agree fairly well in both regions. 
This provides strong evidence that these structures are connected to upstream filaments for these plasma conditions. The rarer observation of elongated filaments in the far-SOL of the lowest divertor view shows that upstream filaments remain connected deep into the divertor in the far-SOL. 
This observation is consistent with the high 
correlation between divertor and midplane fluctuations observed in magnetically connected positions \cite{OffedduCross-fieldTCV}. The present measurements further show that the midplane filaments do not necessarily reach the outer target at all radial locations. 

Indeed, the small scale turbulence in the divertor leg can not be attributed to filaments above the X-point. Mapping their locations and CAS contours along magnetic field lines above the X-point would result in poloidally elongated ribbons with sizes around the ion sound gyro radius. However, no structures with corresponding tilts and shapes are observed above the X-point. 
There is clear evidence that these structures are localised in the divertor and do not exist above the magnetic X-point. This would explain the low correlation of fluctuations along the magnetic field lines between the midplane and the divertor near-SOL, observed in comparable scenarios \cite{OffedduCross-fieldTCV}.

\begin{figure}
        \begin{subfloat}
            \centering
            \includegraphics[clip,width=.5\columnwidth]{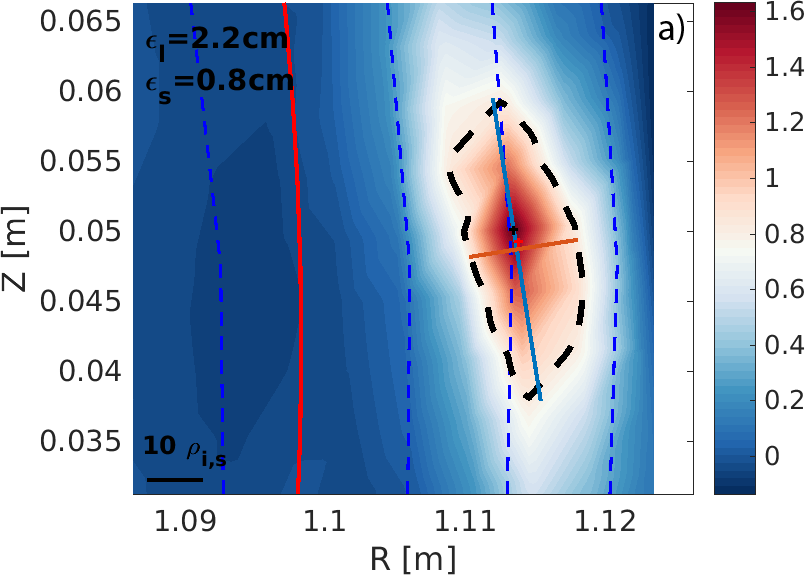}
            \label{fig:mean and std of net14}
        \end{subfloat}
        \begin{subfloat}
            \centering 
            \includegraphics[clip,width=.5\columnwidth]{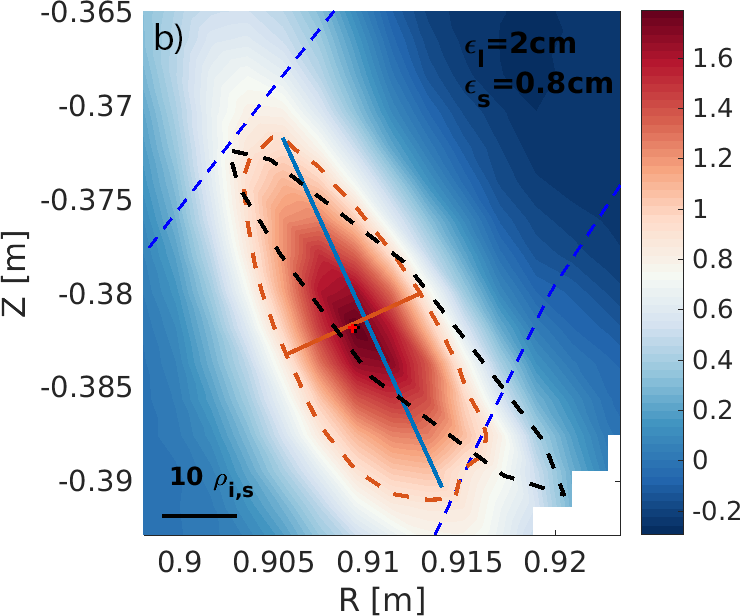}
            \label{fig:mean and std of net24}
        \end{subfloat}
        \newline
        \begin{subfloat}
            \centering 
            \includegraphics[clip,width=.5\columnwidth]{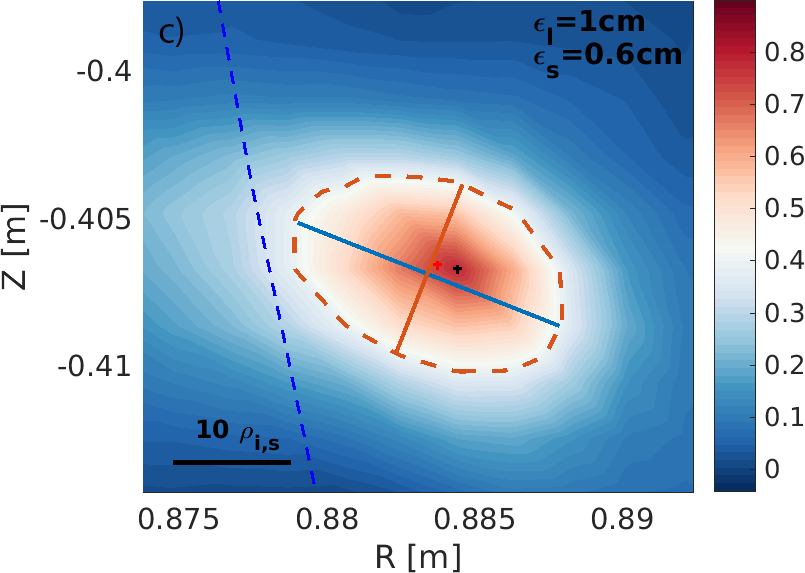}
            \label{fig:mean and std of net34}
        \end{subfloat}
        \begin{subfloat}
            \centering 
            \includegraphics[clip,width=.5\columnwidth]{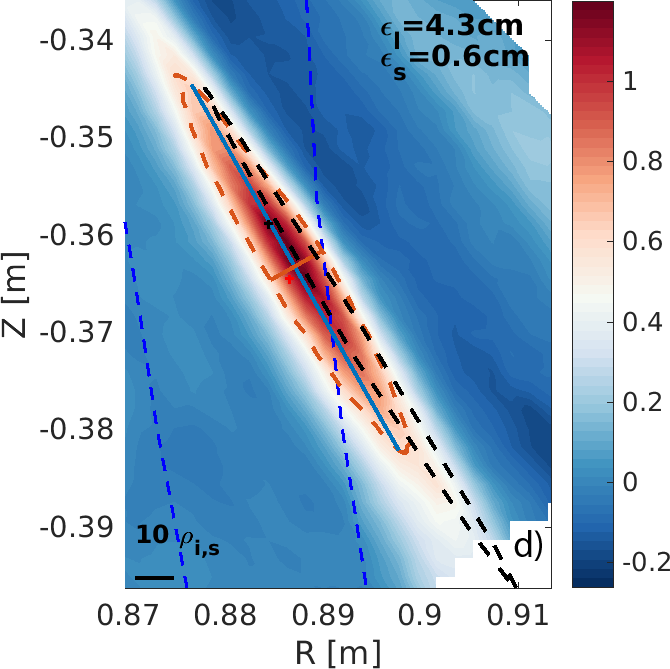}
            \label{fig:mean and std of net44}
        \end{subfloat}
        \caption{Shapes of 2D filaments observed in CAS trigger locations at the outboard midplane a) and above the X-point on $\rho_{\Psi} \approx 1.035$ b). Filaments in the divertor near-SOL c) at $\rho_{\Psi} \approx 1.035$ and in the far-SOL d) $\rho_{\Psi} \approx 1.06$ are alos shown. For each case, the long and short edge size ($\epsilon_l$ blue line and $\epsilon_s$ red line resp.) is specified. The black dashed contour in b) and d) corresponds to the field-line tracing of the midplane contour in a).} 
        \label{fig:casshapes}\vspace{-0.3cm}
    \end{figure}
The filament sizes are quantitatively discussed in fig.\,\ref{fig:casshapes}
, where the sizes of typical filaments are extracted from the CAS frames at typical trigger locations in each region. 
Two characteristic sizes can be taken from the half maximum contour: the longest extent $\epsilon_l$ and the corresponding perpendicular edge $\epsilon_s$ passing through the filaments centre of mass. 
While filaments at the outboard midplane (a) are elongated, approximately, poloidally with $\epsilon_l\!\sim30\,\rho_{s}$, the filaments just above the X-point (b) are elongated mostly radially with a similar $\epsilon_l$. In the far-SOL below the X-point (d), 
the long edge is tilted by magnetic shear, aligning almost with the poloidal direction. 
They 
are less compressed ($\approx\,50\,\%$ larger $\epsilon_s$ and 30\% smaller $\epsilon_l$) than the mapped CAS contours from the outboard midplane (black dashed contour). 
From the toroidal curvature and the LOS misalignment effects, discussed in sec.\,\ref{experiment}, the radial enlargement remains, however, within the diagnostic uncertainty of $\sim5$-$6$\,mm. 
The ion sound gyro-radius $\rho_{s}\,=\,\frac{c_{s}}{\omega_{ci}}=$\,$\frac{\sqrt{2 T_e m_i}}{eB}$\,$\approx0.7$\,mm (assuming $T_e\,\approx\,T_i$\,$\approx\,15$\,eV with ion cyclotron frequency $\omega_{ci}$, ion sound speed $c_{s}$) is shown for scale comparison. 
Typical divertor localised filaments, (c), have dimensions $\sim\,15\rho_{s}$, that are comparable to estimates using line integrated fast imaging on MAST \cite{Harrison2015TheTokamak}.

As shown in fig.\,\ref{fig:velocity}\,a, the filament aspect ratio $\epsilon_l/\epsilon_s\lessgtr3$ provides a good measure to differentiate 
filaments in the divertor. 
$\epsilon_l/\epsilon_s$ increases towards the X-point reaching values of 8 in the far-SOL below the X-point. 
The transition to predominantly more circular, divertor-localised filaments does not appear bound to a flux surface, but rather increases radially beyond a poloidal distance $D_{xpt}\approx 13$\,cm from the X-point. In the lowest observation window, only a few trigger locations at $\rho_{\Psi}\!\geq\!1.06$ show elongated filaments. 
Whether the radial domain governed by divertor localised filaments continues increasing towards the target or remains fixed at $\rho_{\Psi}\approx1.06$ cannot be determined with the present data. 

\subsection{Velocities and turbulent flow pattern}

\begin{figure}[b]
    \begin{picture}(0,200)
         \put(0,0){ \centerline{\includegraphics[width=\linewidth]{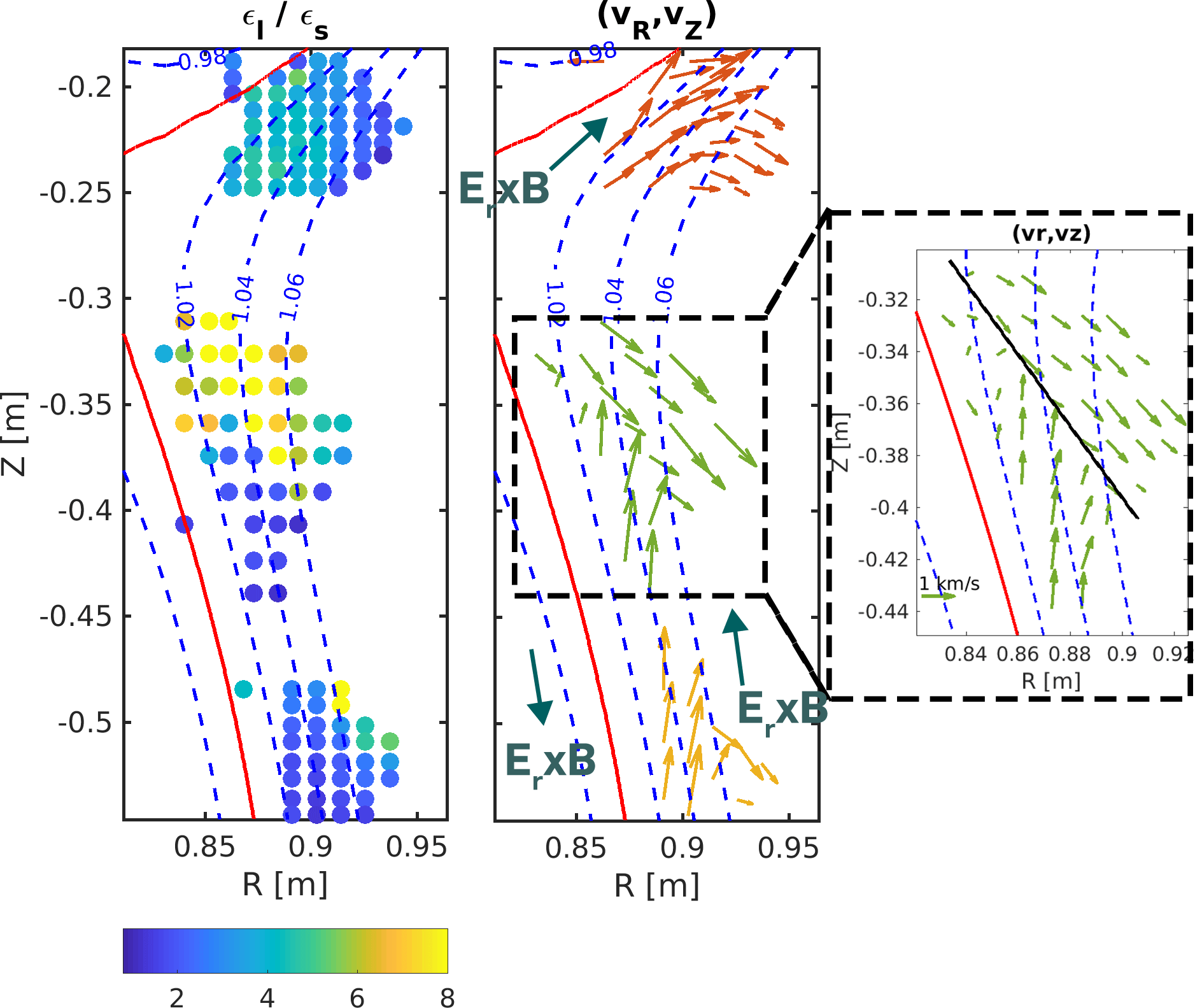}}}
        \put(30,180){\textbf{a)}}\put(105,180){\textbf{b)}}
    \end{picture}
    \caption{a) shows the filament aspect ratio and b) the filament velocity map 
    as deduced form CAS. Only a reduced number of trigger locations is shown, for visibility, with the full resolution for the region below the X-point shown to the right. The black line indicates the transition between the different filament types. The background $\Vec{E}\times\Vec{B}$ direction in the SOL is upwards for these discharges (dark green arrows). \vspace{-0.2cm}}
    \label{fig:velocity}
\end{figure}

\begin{figure}[b]
    \begin{picture}(0,300)
         \put(0,0){ \centerline{\includegraphics[width=.65\linewidth]{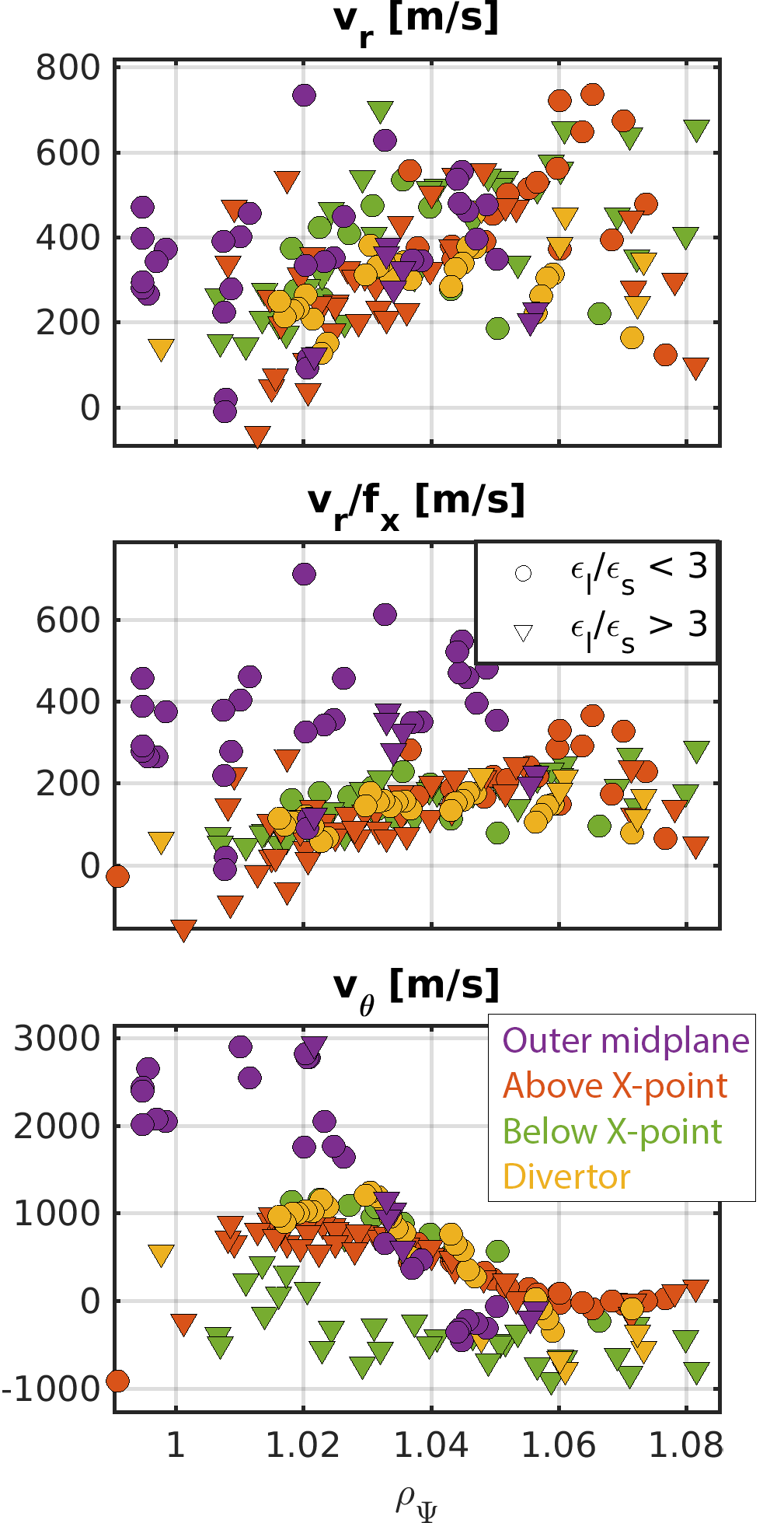}}}
        \put(72,277){\textbf{a)}}\put(72,182){\textbf{b)}}\put(77,86){\textbf{c)}}
    \end{picture}
    \caption{Radial profiles of the radial filament velocity component a), of the radial velocity normalized to the flux expansion $v_r/f_x$ b) and the poloidal velocity c). A separation in symbols is drawn between $\epsilon_l/\epsilon_s\lessgtr3$.} 
    \label{fig:velocity_profile}
\end{figure}
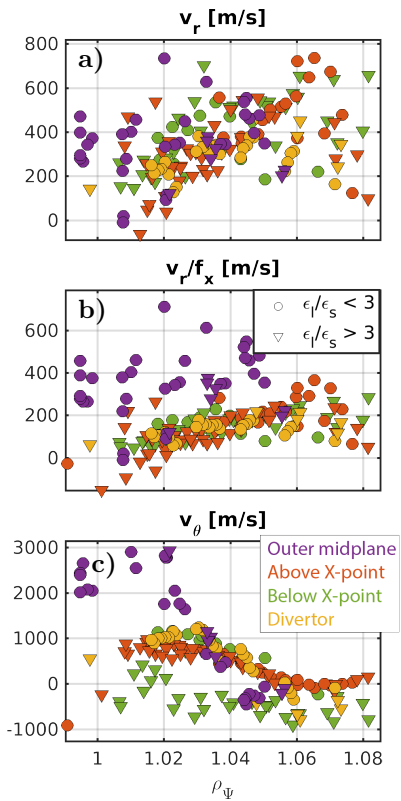

Following the discussion about filament shapes around the X-point, we focus on their cross-field velocities that are in particular important to evaluate contribution to transport. 
Fig.\,\ref{fig:velocity} shows the 2D velocity maps 
illustrating the qualitatively discussed filament displacement seen in fig.\,\ref{fig:snapshots}. 
Below the X-point, both $\epsilon_l/\epsilon_s$ and the velocity map show a clear separation between upstream connected and divertor localised filaments. Due to their opposite poloidal propagation directions, complex interactions can occur at the interface.
The filament velocities are split in perpendicular and tangential components to the flux surfaces, labelled $v_r$ and $v_{\theta}$ respectively, and compared between up- and downstream in fig.\,\ref{fig:velocity_profile}. 
$v_r$ is radially increasing away from the separatrix in all observed regions. 
Velocities typically range from $\approx$\,200\,m/s at $\rho_{\Psi} =1.02$ up to 600\,m/s at $\rho_{\Psi} =1.06$. 

The constant radial velocity up to the LCFS at the outboard midplane indicates that filaments are moving from the core into the SOL. This is less clear in the region just above the X-point, where $v_r$ decreases to zero at the separatrix, suggesting low levels of filamentary 
transport from closed to open field lines in this region.

Despite differences in poloidal sizes 
in the different poloidal regions, $v_r$ 
is largely invariant across most of the SOL. In contrast, C-Mod \cite{Terry2009SpatialAlcator-C-Mod}, observed a 
three-fold higher $v_r$ above the X-point compared to the outboard midplane. As the flux expansion $f_x$ increases strongly from midplane to the X-point, 
$v_r/f_x$ would be constant if the filament remains perfectly field-aligned \cite{Nespoli2020AVelocity}. Fig.\,\ref{fig:velocity_profile}b shows that this is well satisfied, within experimental scatter 
in the data, for all regions around the X-point and divertor. Significantly higher values of $v_r/f_x$ are, however, measured at the outer midplane. This is currently not understood and will need to be further explored. 

The poloidal velocity $v_{\theta}$ (panel c) illustrates the $\epsilon_l/\epsilon_s\lessgtr3$ separation criteria: 
elongated, far-SOL structures display a negative $v_{\theta}\,\sim\!-500$\,m/s in the divertor leg. While this is consistent with the displacement expected from field-line mapping (see fig.\,\ref{fig:snapshots}), it is in the opposite direction to the background poloidal $\Vec{E}\times\Vec{B}$ (pointing upwards for these reversed $B$ discharges). Divertor localised filaments (green and yellow circles) show a positive poloidal velocity, in line with the local $\Vec{E}\times\Vec{B}$ direction. 
Their $v_{\theta}$ dependence is similar to filaments just above the X-point, decreasing from a magnitude of $\approx\,$1\,km/s to near zero around $\rho_{\Psi}\!\approx\!1.05$. Large poloidal velocities, up to $\sim\!3\,$km/s, are measured at the outboard midplane in the local $\Vec{E}\times\Vec{B}$ direction for $\rho_{\Psi}\in[1.0, 1.03]$, presumably due to the steeper gradients at the midplane. This is not observed in the other regions.

\subsection{Single-point measurements and their ability to identify divertor-localised filaments}
So far, we discussed 2D measurements of filaments around the X-point. However, often only single-point measurements are available. We deduce here quantities from single-point measurements that are relevant for the subsequent flux contribution analysis. 
At the same time, we explore whether the presence of 
divertor filaments can be determined single-point parameters. 

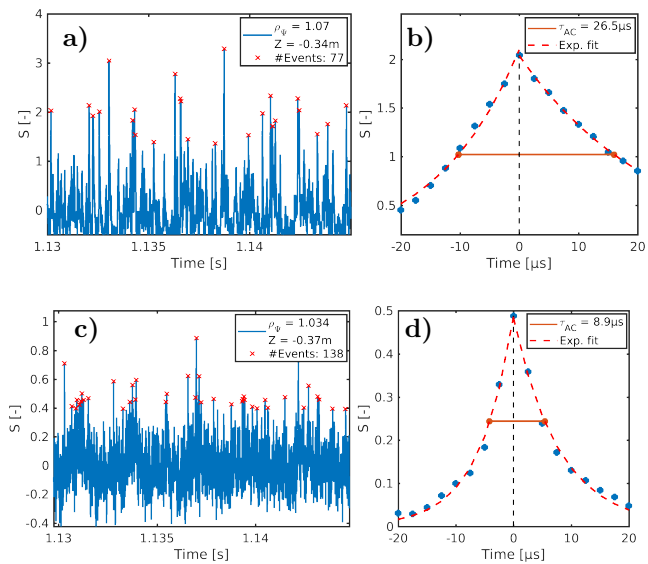
\begin{figure}[b]
        \begin{picture}(0,210)
            \put(0,0){\centering {\includegraphics[clip,width=\columnwidth]{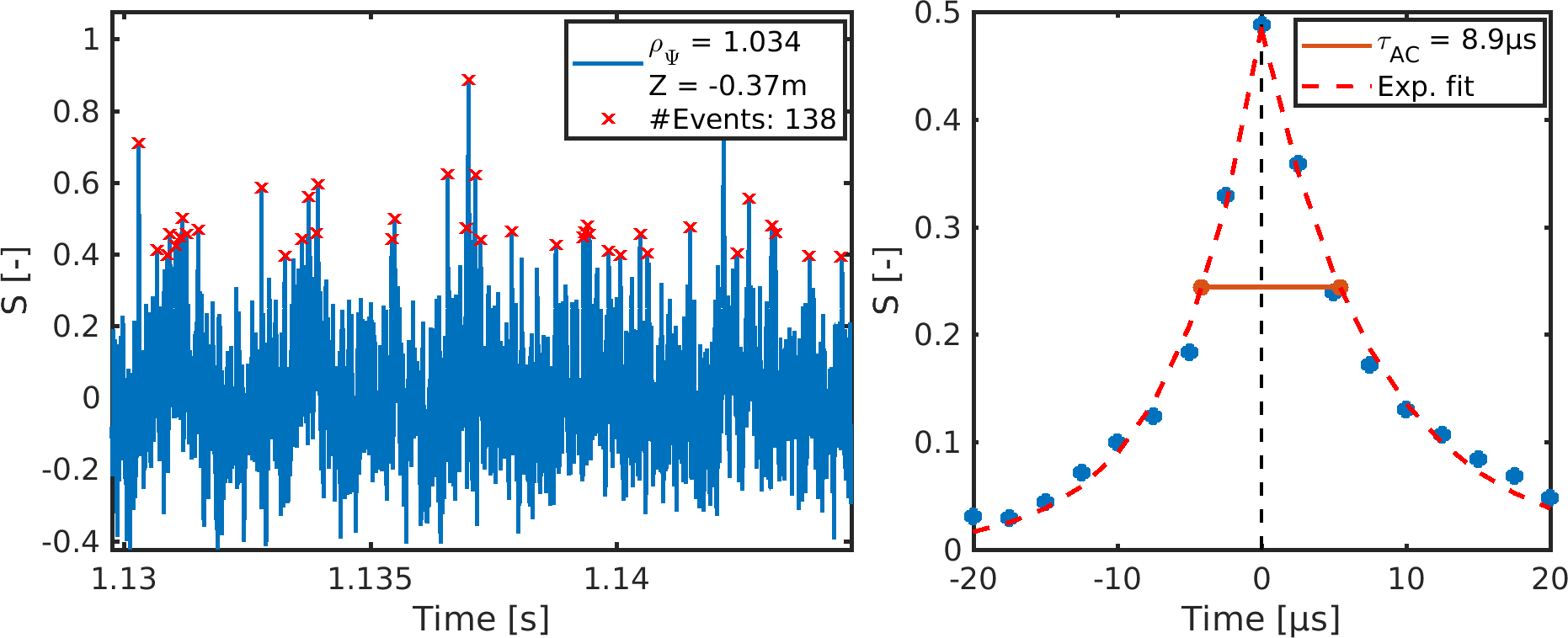}}}\put(20,195){\textbf{a)}}\put(150,195){\textbf{b)}}
        \end{picture}\label{fig:farsoltrace}
        \begin{picture}(0,105)
            \put(0,110){\centering  \includegraphics[clip,width=\columnwidth]{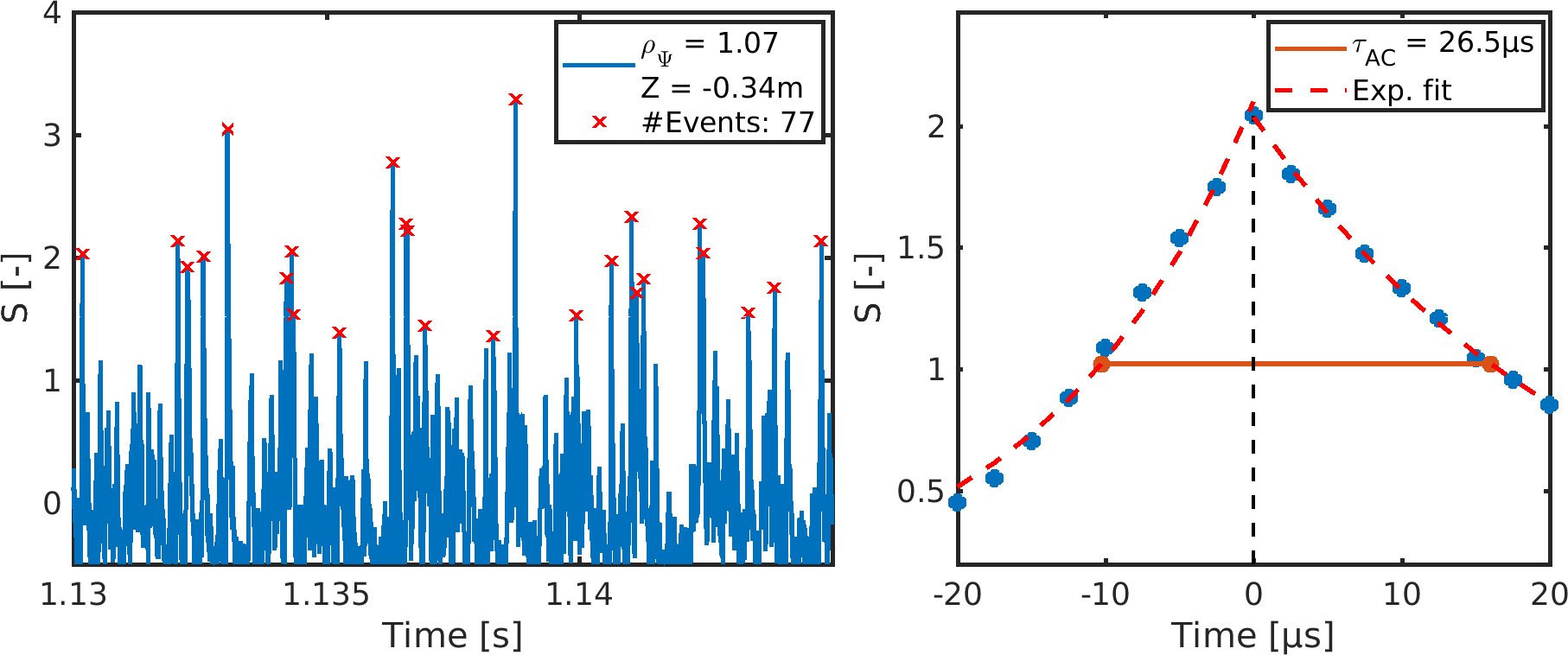}}\put(21,85){\textbf{c)}}\put(146,85){\textbf{d)}}
        \end{picture}\label{fig:nearsoltrace}
        \caption{The normalised signal trace $S$ and the auto-correlation waveform of two trigger locations below the X-point: far-SOL with upstream-connected filaments (a-b) and near-SOL with divertor localised filaments (c-d). The number of events is counted over the 50\,ms of GPI measurement.\vspace{-0.3cm}} 
        \label{fig:soltraces}
    \end{figure}

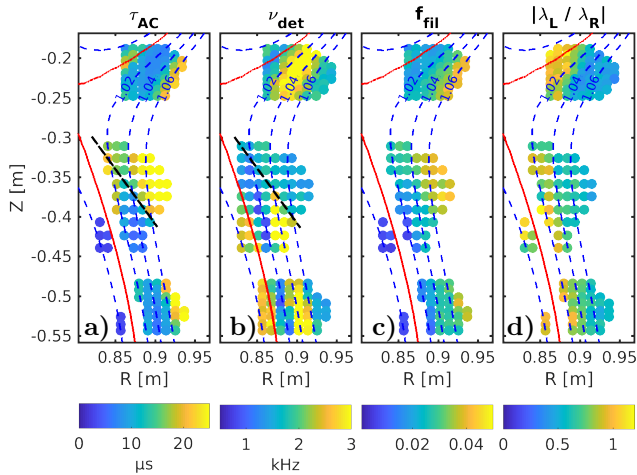
\begin{figure}[b]
\begin{picture}(0,170)
    \put(0,0){\centerline{\includegraphics[width=\linewidth]{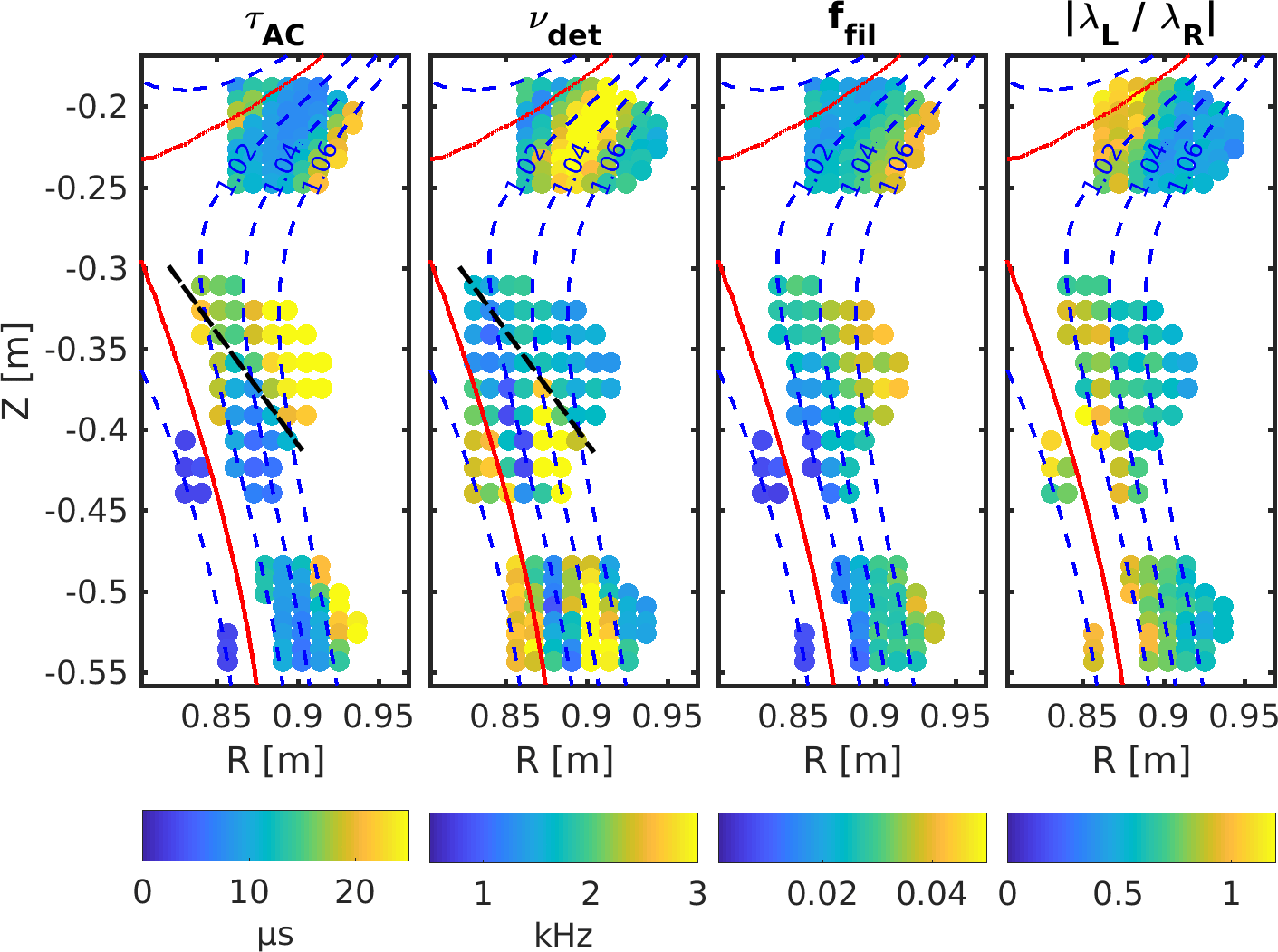}}}\put(28,52){\textbf{a)}}\put(83,52){\textbf{b)}}\put(137,52){\textbf{c)}}\put(188,52){\textbf{d)}}
    \end{picture}
    \caption{Filament auto-correlation time a), filament detection frequency b) and packing fraction c) in the three different SOL regions. Panel d) shows the ratio between the e-folding time of the increasing ($\lambda_L$) and decreasing ($\lambda_R$) part of the auto-conditionally-averaged filament waveforms.}
    \label{fig:properties}
\end{figure}

Fig.\,\ref{fig:soltraces} plots the normalised time traces of two selected pixels 
representative for far-SOL elongated filaments and divertor localised filaments in the region below the X-point (same locations as for fig.\,\ref{fig:casshapes}c and d). Elongated filaments show stronger and fewer peaks compared to divertor localised filaments (red crosses indicate CAS events). The auto-conditional wave-forms are shown in b) and d). Divertor localised filaments show a near symmetric wave-form, whereas elongated filaments show a steeper rise and longer tail, as often observed at the outer midplane \cite{Garcia2006RadialFilaments}. The auto-correlation time $\tau_{AC}$ is determined from the FWHM of these auto-conditional wave-forms. 
The filament packing fraction $f_{fil}$, an important quantity characterising the fraction of time filaments occupy in a specific location, is evaluated as $\nu_{det}\;\tau_{AC}$ \cite{Beadle2020UnderstandingLayer}
, where $\nu_{det}$ is the filament detection frequency. 
These three quantities are shown in fig.\,\ref{fig:properties}.

Fewer upstream-connected filaments, $\nu_{det}\!\sim\!1.5$\,kHz, are detected below the X-point compared to just above the X-point, indicating that not all turbulent structures survive the magnetic shearing into the divertor. 
A higher $\tau_{AC}$ is seen 
below the X-point, in agreement with propagation in the direction of their own major axis. 
Divertor localised filaments have a clear region of appearance with a 3-fold higher $\nu_{det}\!\sim\!$4\,kHz and 
shorter $\tau_{AC}$. 
The transition between the two states is rather sharp, similarly as seen for the velocity in fig.\,\ref{fig:velocity_profile}. 
Low $\nu_{det}\leq0.8$\,kHz measured in both divertor views close to the separatrix shows little fluctuation activity, consistent with qualitative observation in sec.\,\ref{blobshapes} and in line with a quiescent region for $\rho_{\Psi}\!\in\![1.0, 1.02]$. 
The packing fraction increases smoothly radially from the separatrix in all regions with $f_{fil}\!\sim\!2$-5\,\%, comparable to measurements in attached conditions on ASDEX-U \cite{Carralero2018OnContent}. Based on this quantity, no distinction can be made between connected and divertor localised filaments. 
The ratio of increasing and decreasing e-folding time of the filament auto-conditional waveform, which has been used to analyse divertor turbulence on ASDEX-U \cite{Nem2021QuiescentUpgrade}, increases from above to below the X-point at $\rho_{\Psi}\approx1.04$. However, for $\rho_{\Psi} <1.04$ it is hard to discern a clear poloidal difference as well as a separation between filament types in the divertor. 
We conclude that single-point measurements alone can be used to determine the presence of divertor localised filaments in the outer divertor leg, with $\nu_{det}$ and $\tau_{AC}$ the most promising quantities.

\section{Estimates of the role of divertor cross-field filamentary transport}\label{fluxcontribution}
Following identification and description of divertor localised filaments, it becomes pertinent to question whether they contribute to particle and heat flux broadening.
To address this question, we express the filament-induced radial particle flux $\Gamma_{r,fil}$ \cite{Carralero2018OnContent} as
\begin{equation}
    \Gamma_{r,fil} = f_{fil} \; n_{fil} \langle v_r\rangle_{fil}
\end{equation}
with $\langle v_r\rangle_{fil}$ the average radial filament velocity (shown in fig.\,\ref{fig:velocity_profile}a and $n_{e,fil}$ the electron density within the filament. 
The particle flux can, however, not be characterised solely by GPI, 
as it does not directly measure $n_{e,fil}$ 
\cite{Zweben2017InvitedDevices,Scotti2018DivertorNSTX-U}. 
Instead, the importance of the filament radial particle transport 
can be characterised by the ratio of $\Gamma_{r,fil}$ to the parallel convective transport \cite{Carralero2018OnContent}, written in eq.\,\ref{eq:fluxratioinit}. The electron density appears in the ratio of filament density $n_{e,fil}$ to the average density $\langle n_e \rangle$, which can be related to the normalised radiance measured with GPI as follows. 
There is evidence that in the SOL, $n_e$ and $T_e$ fluctuations are strongly correlated \cite{Kube2019StatisticalC-Mod,Wersal2017ImpactDiagnostics}
, such that $\frac{n_{e,fil}}{\cs{n_e}} \approx r \frac{T_{e,fil}}{\cs{T_e}}$, with $r$ a proportionality constant. With this approximation, eq.\,\ref{eq:normalizedbrightapprox} can be solved for $n_{e,fil}/\cs{n_e}$ and substituted into eq.\,\ref{eq:fluxratioinit} to obtain eq.\,\ref{eq:fluxratio}.

\begin{align}
    \frac{\Gamma_{r,fil} }{\Gamma_{||,conv}} &= \frac{n_{e,fil}}{\cs{n_e}} \; \frac{f_{fil} \cs{v_{r,fil}}}{\cs{c_s} \cs{M_{||}}} \label{eq:fluxratioinit}\\
    &\approx \left( \frac{S}{\alpha +r\beta}+1 \right) \frac{f_{fil} \cs{v_{r,fil}}}{\cs{c_s} \cs{M_{||}}} \label{eq:fluxratio}
\end{align}
$\cs{M_{||}}$ is the parallel Mach number and $c_s$ is calculated from time-averaged RDPA and TS 
below and above the X-point, respectively. An approximately constant value of 
$\cs{M_{||}}\!\sim\!0.1$ (towards the target on the LFS) is measured by the RDPA across the SOL in the vicinity of the X-point. 

Several assumptions can be made on $r$. 
For quantitative comparisons, linear machines \cite{Vincent2021High-speedMatters} showed that temperature fluctuations are important in the interpretation of high-speed camera data, especially at low $T_e\sim2$-4\,eV, where $\beta\sim$4$\,\gg\alpha$. At higher $T_e\gtrsim10$\,eV, however, the $T_e$ dependence reduces significantly and $\alpha\sim\beta$ \cite{Zweben2017InvitedDevices}. 
Without available $r$ measurements, 
we consider negligible $T_e$ fluctuations \cite{Carralero2018OnContent}, i.e. $r\approx0$. 
We note that a non-zero, but poloidally uniform $r$, would not change the conclusions hereafter. 
The exponent $\alpha$ is obtained from data provided in \cite{Zweben2017InvitedDevices}, using the local average density $\langle n_e \rangle$.
\begin{figure}[t]
    \begin{picture}(0,230)
         \put(0,0){ \centerline{\includegraphics[width=\linewidth]{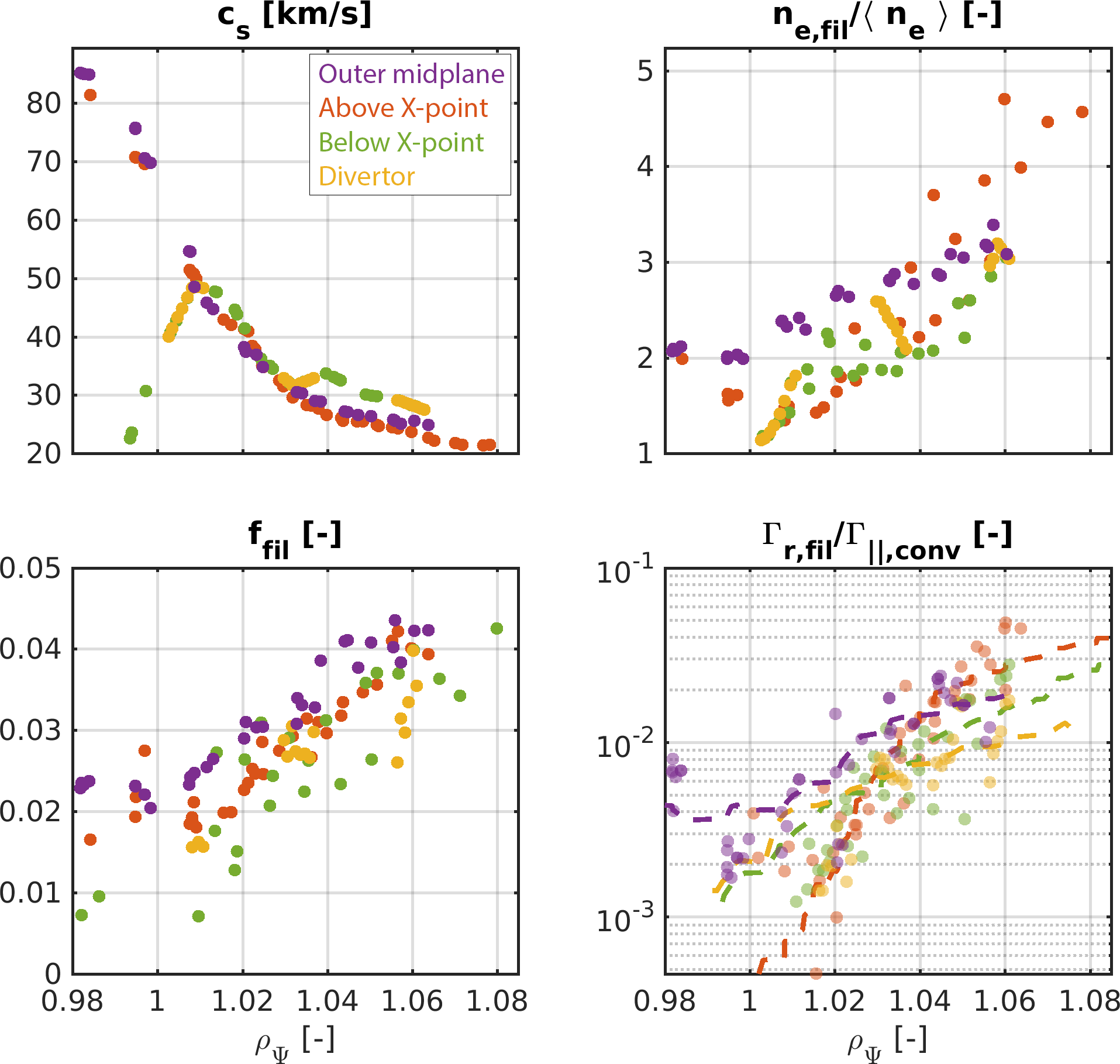}}}
        \put(25,205){\textbf{a)}}\put(148,205){\textbf{b)}}\put(22,93){\textbf{c)}}\put(148,93){\textbf{d)}}
    \end{picture}
    \caption{Radial profiles of ion sound speed a), the GPI deduced quantity related to the density fluctuation amplitude b), the packing fraction $f_{fil}$ c) and the ratio of radial filament induced particle flux and parallel convective flux d) for the probed regions.}
    \label{fig:fluxratio}
\end{figure}
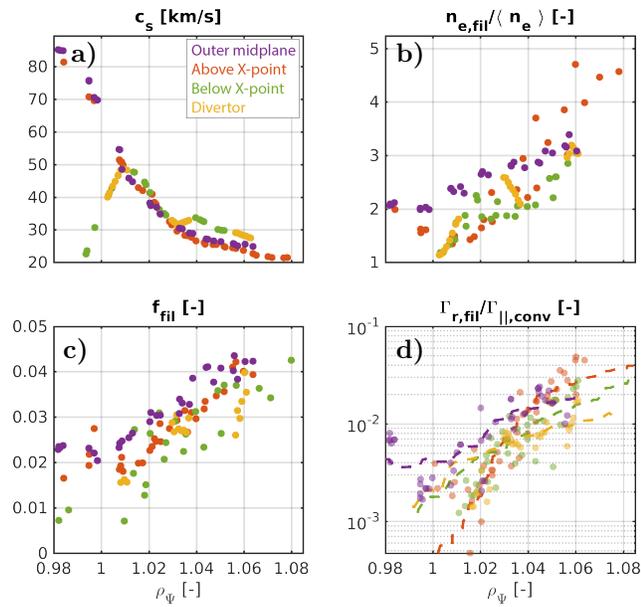

Fig.\,\ref{fig:fluxratio} plots the resulting profiles of the flux ratio (eq.\,\ref{eq:fluxratio}) and the contributing quantities. The ion sound speed decreases with $\rho_{\Psi}$, tracking the average $T_e$ profiles. Conversely, filament related quantities (i.e. 
$\cs{v_{r,fil}}$ shown in fig.\,\ref{fig:velocity_profile}a, $n_{e,fil}/\cs{n_e}$ and $f_{fil}$ in fig.\,\ref{fig:fluxratio}) 
increase radially across all measurement locations. $n_{e,fil}/\cs{n_e}$ is $\approx\,$50\% larger at the midplane compared to other locations for $\rho_{\Psi}<1.05$, consistent with the $\sigma/\mu$ data in fig.\,\ref{fig:statprofile}. 
The ratio of the fluxes (panel d) 
increases from the near- to far-SOL by around an order of magnitude in all regions. More importantly, however, is that the ratio is comparable in the divertor to the outboard midplane. 
This data suggests that divertor filaments (green and yellow curves) contribute in a similar manner to radial transport as upstream filaments
, and thus contribute significantly to target profile broadening. 

\begin{figure}[t]
\begin{picture}(0,175)
         \put(0,0){
    \centerline{\includegraphics[width=\linewidth]{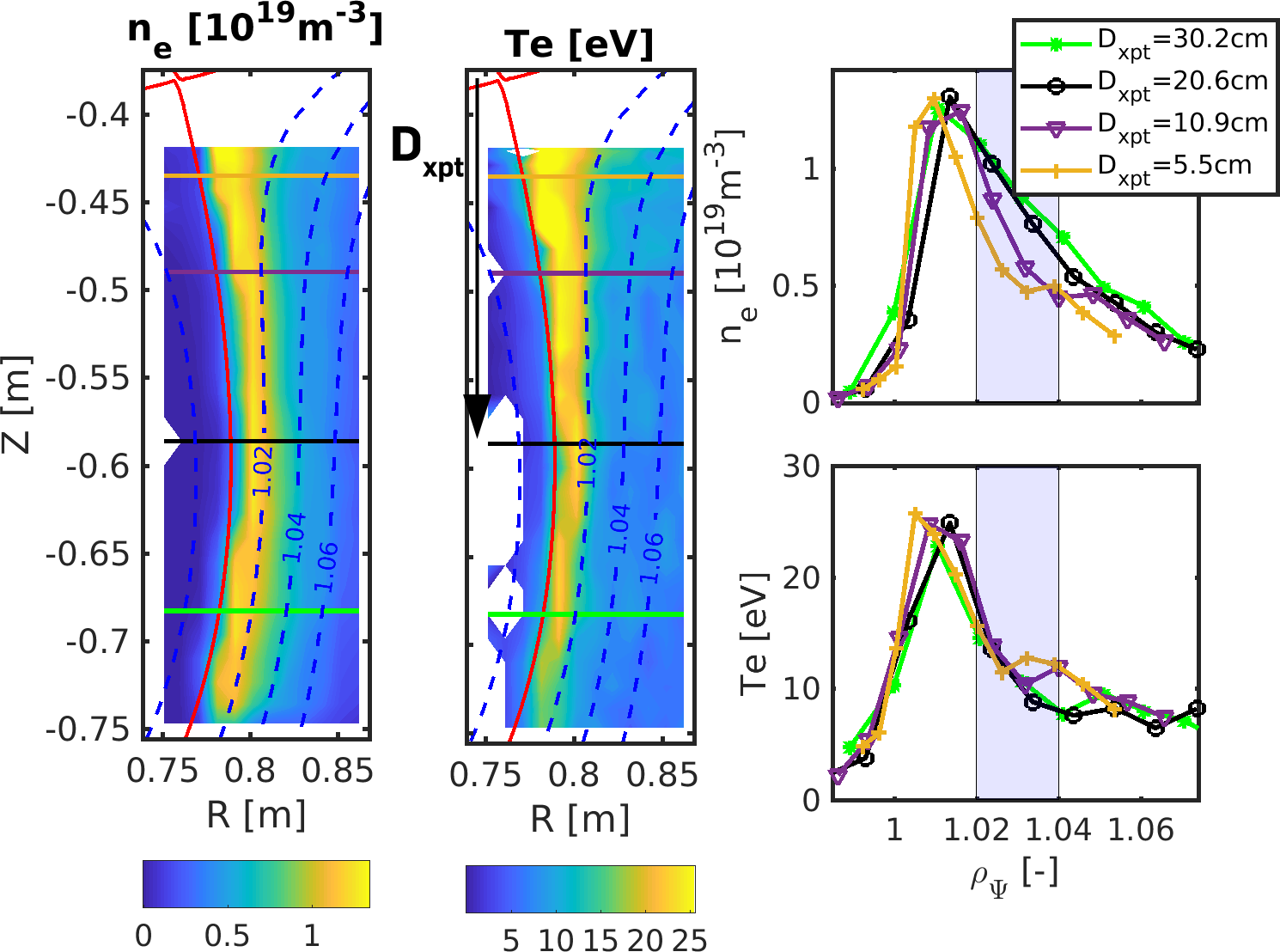}}}\put(60,155){\textbf{a)}}\put(117,155){\textbf{b)}}\put(159,155){\textbf{c)}}\put(159,85){\textbf{d)}}
    \end{picture}
    \caption{2D measurements of electron density a) and $T_e$ b) along the outer divertor leg from RDPA. Radial profiles at 4 distances $D_{xpt}$ to the X-point (colours corresponding to lines on 2D images) are shown in (c-d). The purple shaded area indicates the radial domain where divertor localised filaments are observed.}
    \label{fig:rdpaprof}
\end{figure}
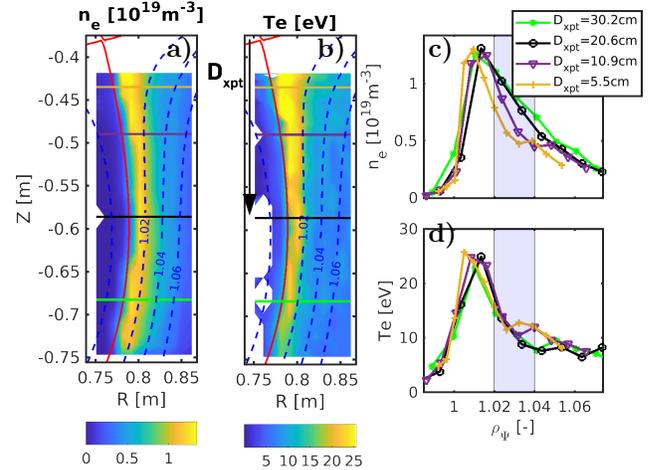
Additional information on divertor profile broadening is obtained from $\cs{n_e}$ and $\cs{T_e}$ measurements from RDPA in 
the red geometry in fig.\,\ref{fig:liuqegeom}, as shown in fig.\,\ref{fig:rdpaprof}\,a)-b). For further insight, the radial profiles at four different vertical positions in panel c)-d). The radial domain governed by divertor localised filaments for $D_{xpt}\!\gtrsim\!15$\,cm is represented by the shaded area. 
The peak values of the profiles show little change along the divertor leg. However, a measurable broadening on the LFS of the peak occurs in $n_e$ as $D_{xpt}$ increases from 5.5\,cm to 20\,cm, from where on no significant further broadening is apparent. 
A comparable broadening is also observed in the ion saturation current. 
These observations support the conclusions based on the preceding estimate of the particle flux contribution that the mechanism 
forming divertor filaments just below the X-point contribute to the broadening. 

Interestingly, the equivalent profiles of $T_e$ do not show this behaviour, suggesting that energy transport behaves differently with respect to the particle transport. It is possible that the filaments mostly contribute to particle transport, without significant temperature variation. 
Measurements were also performed in the other two geometries in fig.\,\ref{fig:liuqegeom}, where $D_{xpt}$ is larger due to the vertical plasma position. Little broadening deeper in the divertor was observed there. 
We mention that the broadening in $n_e$, apparent in fig.\,\ref{fig:rdpaprof}c, could be reproduced in repeat discharges on TCV, also in other baffle configurations, underlying the robustness of this result.

\section{Summary and conclusion}

Taking advantage of the recently installed divertor baffles on the TCV tokamak, a new Gas Puff Imaging diagnostic can observe both the vicinity of the X-point and, by displacing the plasma vertically, the divertor-leg region. 
Images of toroidally localised HeI line emission are acquired at a 400\,kHz frame rate allowing for detailed analysis of SOL turbulence in these regions. 

In this work, experimental investigations in lower single-null, attached L-mode plasmas with reversed magnetic field (B$\times \nabla $B pointing upwards) were performed. The plasma was positioned at three vertical heights, separated by $\pm14$\,cm, to probe above and below the X-point, as well as along the outer divertor leg. 
The results lay out a comprehensive, experimental description of filamentary properties around the X-point and in the outer divertor, which in particular serves as a good basis for the validation of turbulence simulations \cite{Oliveira2022ValidationCase}. 
Elongated filaments were observed both just above the X-point and in the far-SOL in the divertor. Their shape and motion were shown to match well with the expected behaviour based on field line tracing of filaments measured at the outbaord midplane. The reduced occurrence frequency of these filaments when entering the divertor indicates that not all these far-SOL filaments reach into the divertor. 
In the near-SOL ($1.02\!<\!\rho_{\Psi}\!<\!1.06$) and over a large part of the outer divertor, higher frequency divertor localised filaments, fairly circular in shape and with diameters $\sim15\,\rho_{s}$, were observed. Such filaments are generated around the region of peak electron density beyond a certain poloidal distance from the X-point, 
propagating radially outwards at $\sim0.4$\,km/s. 
We observed that their appearance is not bound to a flux surface. Instead, the region they occupy increases radially closer to the outer target, spanning most of the SOL. 
We thus found that filaments move radially outwards at all locations, with divertor-localised filaments having similar $v_r$ as those connected to the midplane. Somewhat surprisingly, $v_r/f_x$ is, however, a factor $\gtrsim2$ larger at the outboard midplane. Poloidal motion is along the local E$\times$B direction, except for upstream connected filaments in the divertor, where field-line connectivity forces them to move in the direction opposite to E$\times$B, away from the X-point. 

Statistical properties of measured SOL fluctuations showed a largely poloidally invariant picture. An exception is at the separatrix and near-SOL at the outboard midplane, where larger relative fluctuation levels and skewness are observed. This, along with positive radial filament velocities, suggests that more turbulence is entering the SOL there. 
The capability to identify divertor localised filaments from single-point measurements alone has also been explored, showing a clear distinction mostly in the occurrence frequency and auto-correlation time. 

The impact of the identified divertor-localised filaments on the radial particle flux is 
estimated to be comparable to that of filaments upstream. 
From in-situ, 2D Langmuir probe measurements along the outer divertor leg, 
some broadening of the radial density profiles was further observed at distances to the X-point where divertor localised filaments start to appear. Hence, they seem to contribute noticeably to the target density and particle flux profile shapes. 
No filaments were observed to move from the peak density location into the PFR and contributing to transport there. 

Further experiments are needed to understand under which conditions divertor localised filaments appear or if upstream filaments can reach the divertor near-SOL in some cases. 
The mechanism causing these divertor localised filaments need to be further explored. 
Experiments are foreseen across a larger plasma parameter space on TCV, including high-confinement regimes. 
The plasma current and collisionality \cite{Nespoli2020AVelocity,Vianello2020Scrape-offRegimes} could, in particular, influence the magnetic and $\Vec{E}\times\Vec{B}$ shear and disconnection of filaments around the X-point along with the magnetic field orientation affecting the drifts in the divertor. 

\section*{Acknowledgements}
The authors thank B. Linehan for valuable discussions. Furthermore, we thank Y. Andrebe and H. Elian for their excellent engineering work on the design of the diagnostic. 
This work was supported in part
by the Swiss National Science Foundation. This work was supported in part by the US Department of Energy under Award Numbers DE-SC0020327 and DE-SC0010529. 
This work has been carried out within the framework of the EUROfusion Consortium,
funded by the European Union via the Euratom Research and Training Programme
(Grant Agreement No 101052200\,-\,EUROfusion). Views and opinions expressed
are however those of the author(s) only and do not necessarily reflect those of the
European Union or the European Commission. Neither the European Union nor the
European Commission can be held responsible for them.
\printbibliography[heading=bibintoc]

\end{document}